\documentclass[MSNbibl,number,citesort,dvips]{arxstspdf}
\usepackage{dcolumn}
\usepackage{graphicx}
\usepackage{flushend}
\usepackage{stfloats}

\volume{27}
\issue{1}
\pubyear{2012}
\firstpage{115}
\lastpage{134}
\doi{10.1214/11-STS363}

\makeatletter
\newcolumntype{d}[1]{D{.}{.}{#1}}
\newcommand{\eqref}[1]{(\ref{#1})}

\newcommand{\given}{\vert}
\newcommand{\ind}{\stackrel{\mathrm{ind}}{\sim}}
\renewcommand{\star}[1]{{#1}^*}
\renewcommand{\bar}[1]{\overline{#1}}
\newcommand{\var}{\operatorname{var}}
\newcommand{\rv}[3][1]{#2_{#1},\ldots,#2_{#3}}
\newcommand{\N}{\mathcal N}
\renewcommand{\a}{\alpha}
\newcommand{\ud}{\,\mathrm{d}}
\newcommand{\abs}[1]{\vert #1 \vert}
\newcommand{\norm}[1]{\Vert #1 \Vert}

\makeatother

\begin{document}
\begin{frontmatter}
\vspace*{6pt}
\title{Shrinkage Estimation in Multilevel Normal~Models}
\runtitle{Shrinkage Estimation}

\begin{aug}
\author{\fnms{Carl N.} \snm{Morris}\corref{}\ead[label=e1,text={morris@ stat.harvard.edu}]{morris@stat.harvard.edu}}
\and
\author{\fnms{Martin} \snm{Lysy}\ead[label=e2,text={lysy@stat.harvard. edu}]{lysy@stat.harvard.edu}}
\runauthor{C. N. Morris and M. Lysy}

\affiliation{Harvard University}

\address{Carl N. Morris is Professor of Statistics, Department of Statistics, Harvard
University, Cambridge,~Massachusetts 02138, USA\\ \printead{e1}.
Martin Lysy is PhD student, Department of Statistics, Harvard University,
Cambridge, Massachusetts 02138, USA\\ \printead{e2}.}
\end{aug}

%
\begin{abstract}
This review traces the evolution of theory that started when Charles
Stein in 1955 [In \textit{Proc.} 3\textit{rd Berkeley Sympos. Math.
Statist. Probab.} \textbf{I} (1956) 197--206, Univ.
California Press] showed that using each separate
sample mean from $k \ge3$ Normal populations to estimate its own
population mean $\mu_i$ can be improved upon uniformly for every
possible $\mu= (\rv\mu k)'$. The dominating estimators, referred to
here as being ``\mbox{Model-I} minimax,'' can be found by shrinking the sample
means toward any constant vector. Admissible minimax shrinkage
estimators were derived by Stein and others as posterior means based on
a random effects model, ``\mbox{Model-II}'' here, wherein the $\mu_i$ values
have their own distributions. Section \ref{eqvar} centers on
Figure \ref{priors}, which organizes a~wide class of priors on the
unknown Level-II hyperparameters that have been proved to yield
admissible \mbox{Model-I} minimax shrinkage estimators in the ``equal variance
case.'' Putting a flat prior on the Level-II variance is unique in
this class for its scale-invariance and for its conjugacy, and it
induces Stein's harmonic prior (SHP) on $\mu_i$.

Component estimators with real data, however, often have substantially
``unequal variances.'' While \mbox{Model-I} minimaxity is achievable in such
cases, this standard requires estimators to have ``reverse
shrinkages,'' as when the large variance component sample means shrink
less (not more) than the more accurate ones. Section \ref{uneqvar}
explains how \mbox{Model-II} provides appropriate shrinkage patterns, and
investigates especially estimators determined exactly or approximately
from the posterior distributions based on the objective priors that
produce \mbox{Model-I} minimaxity in the equal variances case. While
correcting the reversed shrinkage defect, \mbox{Model-II} minimaxity can hold
for every component. In a real example of hospital profiling data, the
SHP prior is shown to provide estimators that are \mbox{Model-II} minimax, and
posterior intervals that have adequate \mbox{Model-II} coverage, that is, both
conditionally on every possible Level-II hyperparameter and for every
individual component $\mu_i$, $i = 1,\ldots, k$.
\end{abstract}

%

\vspace*{-12pt}
\end{frontmatter}

\section{Introduction: Stein and Shrinkage~Estimation}\label{intro}\vspace*{3pt}

Charles Stein \cite{stein55} stunned the statistical world by showing
that estimating $k$ population means $\mu= (\rv\mu k)'$ with their
sample means $y =(y_1,\ldots, y_k)'$ is inadmissible. That result
assumes $k
\ge3$ independent Normal distributions and a sum of mean squa\-red
component errors risk function. With Willard Ja\-mes~\cite{jamesstein61}, he provided a specific shrinkage estimator, the
James--Stein estimator, which dominates the sample mean vector very
substantially.

\begin{figure*}

\includegraphics{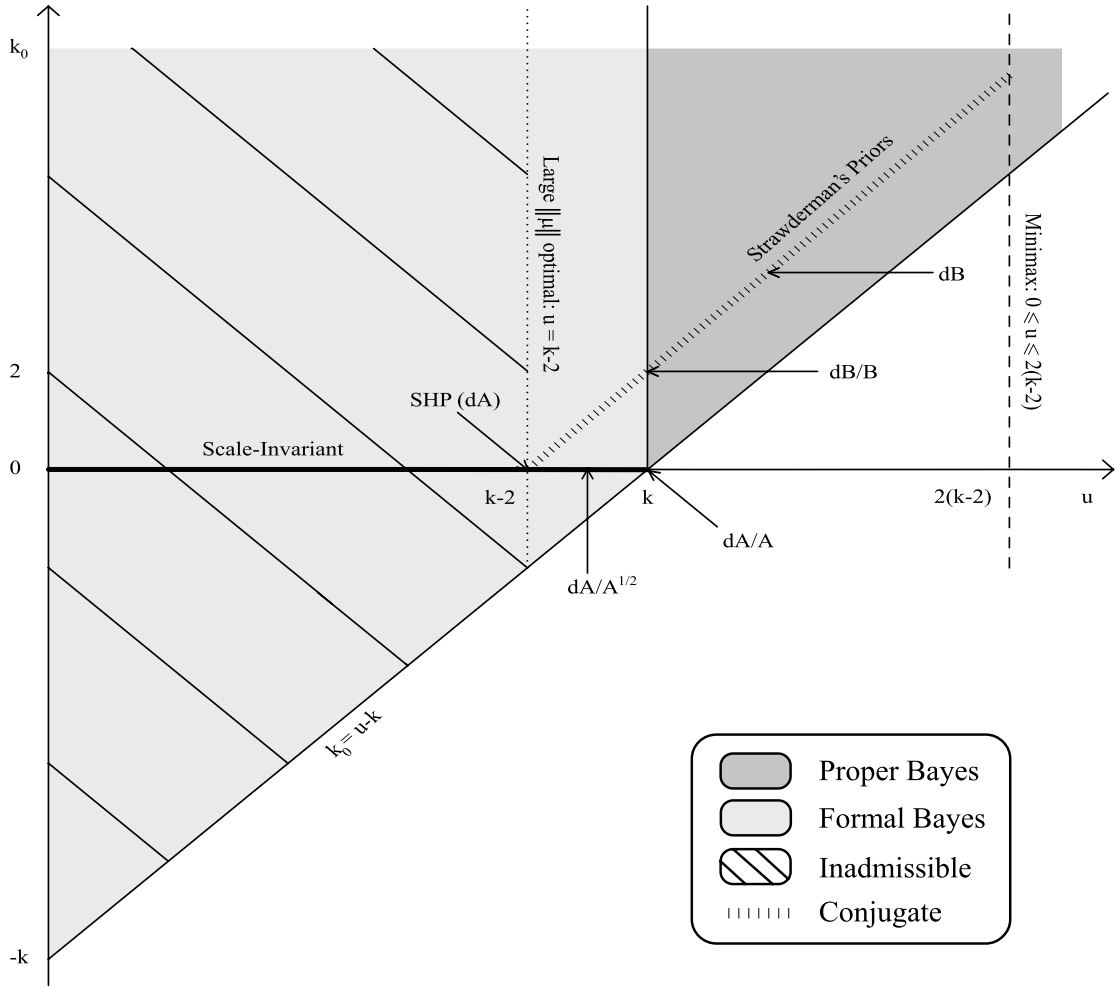}

\caption{Classification of the proper and formal priors of the
form given in \protect\eqref{betaprior}. The $u$-axis determines
limits on
minimaxity, with smaller values providing less shrinkage.
Larger $k_0$ indicates more prior information, and thus more
shrinkage.}\label{priors}
\end{figure*}

This first section introduces the history of the James--Stein minimax
estimator and its extensions when ``equal variances'' prevail, with
\mbox{``Model-I''} evaluations that are conditional on $\mu$. However,
``Mo-\break del-I'' does not allow certain practical needs to be met, such as
valid confidence intervals. Section \ref{eqvar} shows how this has
been rectified by enlarging ``Mo-break del-I'' to \mbox{``Model-II''} wherein random
effects distributions are assigned in Level-II. The resulting
framework enables repeated sampling (frequency based) interval
estimates \cite{efronmorris75,morris83} and frees practitioners from
determining and specifying valid relative weights for each
squared-error component loss, upon which \mbox{Model-I} minimax estimators
depend critically. \mbox{Model-II} even supports developing admissible
minimax shrinkage estimators via posterior mean calculations by
simplifying the specification of prior distributions, proper and
otherwise, on the Level-II parameters. The centerpiece of Section
\ref{eqvar} is Figure \ref{priors}, which graphically organizes some
priors on the Level-II variance that lead to minimax estimators.
Stein's harmonic prior (SHP) in Figure \ref{priors} corresponds to an
admissible shrinkage estimator that provides acceptable frequency
coverage intervals in \mbox{Model-II} evaluations.

Section \ref{uneqvar} reviews the unequal variances case that arises
regularly in practice, but for which mathematical evaluations are
difficult. The previous sections are meant especially to provide the
background needed for more research on the operating characteristics in
repeated sampling of unequal variances procedures, while Section
\ref{uneqvar} shows why that is needed. It is shown in Section
\ref{uneqvar} why, in substantially unequal variances settings, the
\mbox{Model-II} random effects framework works well while the
\mbox{Model-I}
perspective provides inappropriate (``reversed'') shrinkage patterns.
The SHP prior in that setting leads to estimators and to formal
posterior intervals for $\mu_i$, $i = 1, \ldots, k$, that appear to
provide approximate (or conservative) frequency\vadjust{\goodbreak} confidence intervals
with respect to \mbox{Model-II} evaluation standards that for each individual
$\mu_i$ approximates or exceeds its nominal $95\%$ coverage, no matter
what the true Level-II variance.

%
\begin{table}
\caption{Hospital profiling data and James--Stein shrinkage estimates
for $k = 10$ NY hospitals}\label{hospeq}
\begin{tabular*}{\columnwidth}{@{\extracolsep{\fill
}}d{2.0}d{2.2}cccd{2.2}@{}}
\hline
\multicolumn{1}{l}{$\bolds{i}$} & \multicolumn{1}{c}{$\bolds{y_i}$}
& \multicolumn{1}{c}{$\mathrm{\mathbf{sd}}_{\bolds{i}}$} &
\multicolumn
{1}{c}{$\bolds{V_i}$}
& \multicolumn{1}{c}{$\bolds{\hat B_{\mathrm{JS}}}$} & \multicolumn
{1}{c@{}}{$\bolds{\hat\mu_{\mathrm{JS},i}}$} \\
\hline
1 & -2.15 & 1.0 & 1.0 & 0.688 & -0.67 \\
2 & -0.34 & 1.0 & 1.0 & 0.688 & -0.11 \\
3 & -0.08 & 1.0 & 1.0 & 0.688 & -0.02 \\
4 & 0.01 & 1.0 & 1.0 & 0.688 & 0.00 \\
5 & 0.08 & 1.0 & 1.0 & 0.688 & 0.02 \\
6 & 0.57 & 1.0 & 1.0 & 0.688 & 0.18 \\
7 & 0.61 & 1.0 & 1.0 & 0.688 & 0.19 \\
8 & 0.86 & 1.0 & 1.0 & 0.688 & 0.27 \\
9 & 1.11 & 1.0 & 1.0 & 0.688 & 0.35 \\
10 & 2.05 & 1.0 & 1.0 & 0.688 & 0.64 \\
\hline
\end{tabular*}\vspace*{2pt}
\end{table}

%
\begin{figure*}[b]

\includegraphics{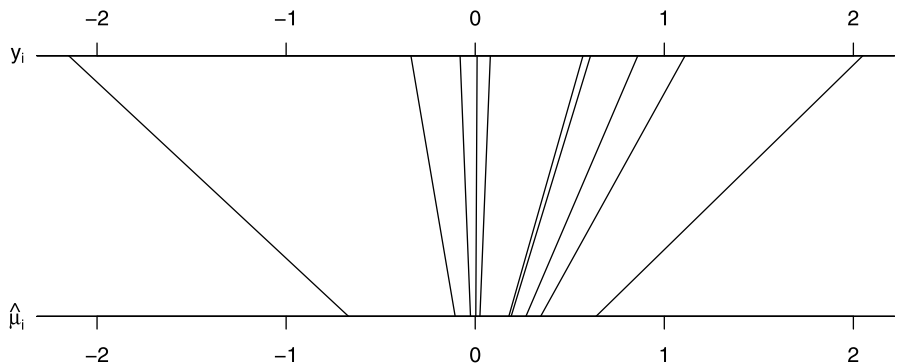}

\caption{Unbiased (top) versus James--Stein (bottom) estimates for 10
NY hospitals.}\label{crosseq0}
\end{figure*}

The data in Table~\ref{hospeq} provide an equal variance example based
on a 1992 medical profiling evaluation of $k = 10$ New York hospitals.
We are to consider these as Normally-distributed indices of successful
outcome rates for patients at these 10 hospitals following coronary
artery bypass graft (CABG) surgeries. The indices are centered so that
the New York statewide average outcome over all hospitals lies near 0.
Larger estimates $y_i$ indicate hospitals that performed better for
these surgeries. For example, Hospital 10 was more than 2 standard
deviations above the statewide mean. All 10 sample means have nearly
the same variances, which we have scaled so the common variance is
about $V = 1.00$. The variances $V_i$ must be the same in order to
meet the equal variance assumption upon which the James--Stein estimator
is based. This ``equal variance'' case enables various mathematical
calculations that are difficult, if not impossible, for the widely
encountered ``unequal variances'' situation.

The vector of sample means $y$ has total mean squared error (risk) as
an estimator of $\mu$ given by
\[
E \sum_{i=1}^k(y_i - \mu_i)^2 = \sum_{i=1}^k V_i = k V.
\]
This unbiased estimator is minimax, since its constant risk is the
limit of the risks of a sequence of proper Bayes' rules (see, e.g.,
Theorem 18 of Chapter 5 in \cite{berger85}).

In the simplest situation, the James--Stein estimator ``shrinks'' $y_i$
toward an arbitrarily preassigned constant $\mu_0$. It is appropriate
to set $\mu_0 = 0$ in this case because we have recentered the CABG
indices to have NY statewide mean equal to 0. Then with $\mu_0 = 0$,
the sum of squared residuals for these data,
\[
S = \sum_{i=1}^{10} y_i^2/V = 11.62,
\]
would have a $\chi^2_{(10)}$ distribution if the hypothesis that all
values of $\mu_i \equiv\mu_0 = 0$ were true, thereby failing to reject
the null at even the 30\% level. However, most members of the medical
community would not believe that all hospitals are equally effective,
and many in the statistical community would be reluctant to think that
the first and last hospitals in the list, whose quality estimates
differ by more than 2 standard deviations from 0, should be declared to
have the same underlying quality as all the others.

On the other hand, $S$ isn't far from its expectation $k = 10$ if all
the $\mu_i$ are 0, and some extreme rates would occur, at least in
part, because of randomness. Thus, regression-toward-the-mean (RTTM),
that is, shrinkage toward $\mu_0$, would be expected if more data were
to appear for these hospitals.

RTTM is anticipated if one believes that there is some similarity among
the hospitals, and that sampling variation is part of the reason for
the extreme hospitals. That is, the hospital with the highest quality
index with $y_{10} = 2.05$ probably has a true mean $\mu_{10}$ smaller
than $2.05$ because
\[
E\Bigl[\max_{1\le i\le k}y_i \vert\mu\Bigr] > \max_{1\le i \le k}E[y_i
\vert
\mu_i] = \max_{1\le i \le k} \mu_i \ge\mu_{10}
\]
(by Jensen's inequality and convexity of the maximum function). So we
expect in this case that the~ob\-served maximum $y_{10} = 2.05$ exceeds
$\mu_{10}$, and a~shrun\-ken estimator is in order. The two-level
\mbox{Model-II}, soon to be described, anticipates and models RTTM, leading to
shrinkage estimation.\vadjust{\goodbreak}

Following earlier notation set in a series of papers by Efron and
Morris, for example, \cite{efronmorris75}, about Stein's estimator and
its generalizations, we denote shrinkage factors by the letter $B$
(often with subscripts). The James--Stein shrinkage coefficient for
this setting is calculated as
\[
\hat B_{\mathrm{JS}} = (k-2)/S,
\]
which for these data is $\hat B_{\mathrm{JS}} = 8/11.62 = 0.688$. 
This estimator then shrinks the usual unbiased estima\-tes~$y_i$ toward
$\mu_0 = 0$ according to
\[
\hat\mu_{\mathrm{JS},i} = (1 - \hat B_{\mathrm{JS}})y_i + \hat B_{\mathrm{JS}}
\mu_0 = (1 - \hat
B_{\mathrm{JS}})y_i.
\]
Based on this shrinkage estimate, future observations are being
predicted to regress about $68.8\%$ of the way toward 0. Column 5 of
Table \ref{hospeq} lists the shrunken values
\[
(1-0.688)\times y_i + 0.688 \times0 = 0.312 \times y_i
\]
for each hospital, the James--Stein estimate of the mean. For example,
the estimate of Hospital 10's quality index is reduced from 2.05
standard deviations above the New York mean to 0.64 standard
deviations. The RTTM effect is strong for these 10 hospitals, which
are estimated to be more similar than different, with only 31.2\% of
the weight allocated to each hospital's own estimate.
Figure \ref{crosseq0} illustrates the shrinkage pattern.

The parameter $\mu_i$ can be thought of as the quality index that would
result for hospital $i$ if that hospital theoretically could have
performed a huge number of CABG surgeries in 1992. Whether the JS
estimator of quality is a better estimator of $\mu$ than $y$ for these
data cannot be guaranteed because the true values of $\mu$ aren't
known. However, one can calculate an unbiased estimator of the
expected risk (i.e., for sum of squared errors) of the JS
estimator\vadjust{\goodbreak}
\cite{jamesstein61}. This unbiased estimator of the risk is
\[
\hat R = V\bigl(k-(k-2)\hat B_{\mathrm{JS}}\bigr),
\]
a function of $y$ only through $S$. That $y$ is inadmissible and that
the JS estimate is ``minimax'' (risk never exceeds $kV$) follows
because $\hat B_{\mathrm{JS}} > 0$ for all data sets. This proves minimaxity,
that the risk of $\hat\mu_{\mathrm{JS}} = (\hat\mu_{\mathrm{JS}, 1}, \ldots,
\hat
\mu_{\mathrm{JS}, k})'$ as a function of $\mu$ is
\[
E[\hat R \vert\mu] = V\bigl(k-(k-2)E\hat B_{\mathrm{JS}}\bigr) < kV.
\]

For these data, $\hat R = 1.00 \times(10 - 8 \times0.688) = 4.496$.
This is a large reduction in mean squared error, less than half of $k V
= 10$, the risk of the separate unshrunken estimates $y_i$. In fact,
the smallest possible value of the risk for the JS estimator is $2V$,
when $\mu= 0$, for any value of $k \ge3$, thus offering very
substantial possible improvements on $y$.

The JS estimator can be extended in the equal variance setting to cover
more general situations. For example, as Stein and others showed, one
can shrink the $y_i$ toward the grand mean of the data, $\bar y = \sum
y_i/k$. With these hospital data this would shrink toward $\bar y =
0.272$ (which differs by less than one standard error of the overall
average for the 10 hospitals from the assumed mean 0). More generally,
if along with each $y_i$ one collects a vector of $r > 0$ covariate
vectors $x_i$ (possibly including the intercept), each $y_i$ can be
shrunk toward its regression prediction $x_i'b$, where
\[
b = (X'X)^{-1}X'y
\]
and $X$ is the $k\times r$ covariate matrix with columns $x_i$, $i = 1,
\ldots, k$. Doing this forfeits $r$ degrees of freedom, so that
\[
\hat B_{\mathrm{JS}} = (k-r-2)/S,
\]
with $S$ now replaced by
\[
S = \sum_{i=1}^k (y_i - x_i'b)^2/V.
\]
The James--Stein estimates of the $\mu_i$ then become
\begin{eqnarray*}
\hat\mu_{\mathrm{JS}, i} & = & (1-\hat B_{\mathrm{JS}})y_i + \hat B_{\mathrm
{JS}} x_i'b \\
& = & (1-\hat B_{\mathrm{JS}})(y_i - x_i'b) + x_i'b.
\end{eqnarray*}
Writing $\hat\mu_{\mathrm{JS}, i}$ this way suggests that shrinking with $r >
0$ does not affect the $r$-dimensional regression space, but only
shrinks toward 0 in the $k-r$ dimensional space orthogonal to it.
Indeed, the problem can be ``rotated'' to an equivalent one in which
the last $r$ values of the residuals $y_i - x_i'b$ are all equal to 0,
regardless of the value of $y$, for example, Stein~\cite{stein66}. The
example just considered, with shrinkage toward zero, shows what happens
to the residuals when shrinkage is toward a regression model.

Of course $V$ needn't be $1.00$, or even be known, provided there
exists an independent Chi-square estimate of $V$. While that can be
handled straightforwardly in the equal variance case \cite{stein66},
it will not be a central issue in any case if the degrees of freedom
are substantial.

Using the JS estimator seems easy and powerful, but many complicating
issues arise in practice:

\begin{enumerate}
\item What is the standard error of each individual estimate?
One hopes the JS estimator for Hospital~10 improves $y_{10} = 2.05$
(with standard devia\-tion${} = 1.00$) by using the better
estimate $\hat\mu_{10} = 0.64$. The sum of individual risks has
decreased from 10 to 4.5 for all 10 hospitals, but this does not
mean the variance for each individual estimate has dropped to 0.45.
Furthermore, the JS estimator cannot even guarantee that every
component (hospital) has a~smaller risk (expected squared error)
as a function of $\mu$. Such an improvement is impossible because
each individual $y_i$ is an admissible estimate of its own $\mu_i$,
in one dimension. Rather, minimaxity of the JS estimator for sum
of squared errors is accomplished by ``balancing'' or ``trading off''
component risks. Components with mean square errors that exceed
$V$ are guaranteed to have their risks more than offset by risk
improvements on the remaining components. The minimaxity claim
(improvement on the unshrunken vector of unbiased estimates) is
for aggregate risk, and not for every component.

\item Why, even in this equal variance case, should the loss function
be an unweighted sum of squares? In applications the loss function
could require different relative weights to reflect unequal economic
loss for the mean squared errors of different components
(hospitals, here). That is, the appropriate loss function could be
\[
L(\hat\mu, \mu) = \sum_{i=1}^k W_i(\mu_i - \hat\mu_i)^2
\]
for some appropriate weights $\rv W k > 0$.

Users of the James--Stein estimator typically assume that all $W_i$
are equal in assessing its risk benefits. But would NY hospital
administrators agree that hospital errors can be traded off with
equal weights? Perhaps weights should differ for teaching hospitals,
or for military hospitals, or for children's or other specialty
hospitals, or for hospitals in areas far from medical centers, or
for large hospitals. Getting agreement on that issue has arisen
with various real shrinkage applications. Even if the administrators
could agree on the values of the $W_i$, the James--Stein estimator
would not dominate $y$ when the $W_i$ are sufficiently unequal.
There is a way out that seems reassuring, at first, because a
shrinkage estimator can be found to dominate~$y$ for any given
weights $W_i$. But there is a rub. The dominating estimator for
a set of weights depends on the specified weights, and then it
cannot be expected to dominate~$y$ for a different set of weights.
Only the unshrunken estimator $y$ can be guaranteed to be minimax
independently of the weights $W_i$. Its risk, the minimax risk,
is $V = \sum W_i$. More on this in Section~\ref{uneqvar}.

\item Even with equal weights, $W_i\,{\equiv}\,1$, another \mbox{problem}
arising in practice and in the theory is that~$\hat B_{\mathrm{JS}}$
can exceed 1. A (uniformly) better shrinkage constant uses
$\min(1, \hat B_{\mathrm{JS}})$ instead and easily is seen to
reduce the total risk. That change necessitates developing a~new unbiased estimator of risk. This was made possible, and
easy, by a~simple calculus pioneered by Stein \cite{stein73,stein81}
and independently by Berger's integration by parts technique \cite{berger76}.

This truncated shrinkage estimator's improvement shows that the
James--Stein estimator is inadmissible itself. The improved
truncated estimator also is inadmissible, as it has a discontinuous
derivative, while admissible estimators must have all their derivatives
(as a function of the data). The search for admissible estimators
began soon after the James--Stein estimator, for example,
Stein \cite{jamesstein61} and Brown \cite{brown66}.

\item We already have noted that there is no agreed-upon way
to estimate the component variances of the JS estimator.
Correspondingly, there is no way to determine separate confidence
intervals for each $\mu_i$. Confidence ellipsoids, for example,
Stein \cite{stein62} and Brown \cite{brown09}, can be and have
been developed for the equal variance setting. However, ellipsoids
may be unattractive to a data analyst who has the alternative
of estimating with~$y_i$ and using $V^{1/2}$ as the standard
error, with a~corresponding exact confidence interval for
each component obtained via the Normal distribution.
Unfortunately, only aggregates (ellipsoidal sets in this context)
can provide uniformly better coverage if coverage must hold
conditionally on the underlying $\mu$ for all $\mu$, that is,
with \mbox{Model-I} evaluations. There is no agreed upon
component-wise procedure for standard errors and intervals
for individual components $\mu_i$ simply because no such
procedure is possible as a function of $\mu$. This problem
(and others too) can be rectified only via acceptance of a
two-level, random effects model referred to here as \mbox{Model-II}.

\item The overriding difficulty for the JS estimator as
a~practical tool for data analysts is that, except for data
produced by carefully designed experiments, real data rarely
occur with equal variances $V_i = V$. Even the hospital data of
Table~\ref{hospeq} do not have exactly the same variances. The
first author has participated in developing and in using shrinkage
techniques for hospital profiling and for other applications
(e.g., \cite{morris83,christiansenmorris97}) without ever seeing
hospital or medical data with equal variances, simply because
hospital caseloads (numbers of patients) vary considerably.
For this initial discussion to illustrate the JS estimator and
related shrinkage procedures in the equal variances setting, we
have picked 10 of the 31 hospitals (the 31 to be described later)
that had similar variances. These 10 each have sample sizes
within 15\% of 550 patients.
\end{enumerate}

%
%


\section{Theoretical and Bayesian Developments for the Equal
Variance~Case}\label{eqvar}

This section reviews expansion of the assumptions of ``\mbox{Model-I}'' to a
two-level model, ``\mbox{Model-II},'' which at Level-II includes a random
effects model on the~$\mu_i$, with the Level-II parameters unknown but
estimable from the data. \mbox{Model-II} and Stein's harmonic prior (SHP), to
be introduced in this section, will be especially important as a basis
for developing frequency procedures in the difficult unequal variances
situation of Section \ref{uneqvar}. After briefly introducing the
unequal variances case in this section, the equal variances setting is
studied because of its relatively easy calculations. This enables
Bayesian analysis that uses formal priors on the Level-II parameters
that produce shrinkage estimators as posterior means. In the equal
variance setting, many of these estimators have been proven to be
minimax (some also are admissible) in the original \mbox{Model-I} sense of
Stein, that is, for total square error loss and for every possible mean
vector $\mu$. The centerpiece of this section is
Figure~\ref{priors},\vadjust{\goodbreak}
which displays graphically certain famous distributions on the Level-II
variance, ``$A$'' that are known to provide minimax shrinkage
estimators. Of central importance is Stein's harmonic prior (SHP) on
$\mu$, which stems from imposing an improper flat prior on $A$ and
yields an admissible, minimax modification of the James--Stein
estimator. This SHP shrinkage estimator leads to posterior interval
estimates that meet confidence requirements for coverages in \mbox{Model-II}
evaluations.

A generalization of the James--Stein estimator almost always is required
in practice because the unequal variance situation arises, and also
because data analysts often must provide interval estimates. Uniform
risk dominance as a function of $\mu$ will be seen in Section
\ref{uneqvar} to require inappropriate (reversed) shrinkage patterns in
practice. Of course, shrinkage methods are used commonly in
applications, almost always being based on a two-level random effects
mo\-del, the~$\mu_i$ being random effects with their own distributions.
Such models belong to frequentists and Bayesians alike, known as
hierarchical models, multilevel models, empirical Bayes models and by
other terms. Table~\ref{model} shows one such model with Normally
distributed observations (Level-I), and Normally distributed random
effects (Level-II). The two columns, that is, the Descriptive and the
Inferential versions of the model, are equivalent in that both sides
give rise to the same joint distributions of the data and the random
effects, $(y, \mu)$, given the hyperparameter $\alpha$ that governs the
joint distribution. These models allow ``unequal variances'' $V_i$,
perhaps because $V_i = \sigma^2 / n_i$ with different sample sizes.
That anticipates Section \ref{uneqvar}, but in this ``equal variances''
Section we always assume $V_i \equiv V$.

%
\begin{table*}
\tablewidth=350pt
\caption{Multilevel model layout}\label{model}
\begin{tabular*}{350pt}{@{\extracolsep{\fill}}lll@{}}
\hline
\textbf{Level} & \textbf{Descriptive version} & \textbf{Inferential
version}\\
\hline
\phantom{II}I & $y_i \vert\mu_i \ind\N(\mu_i, V_i), \ i = 1,
\ldots,
k$ & $y_i \vert\a\ind\N(x_i'\beta, V_i + A)$ \\
\phantom{I}II & $\mu_i \vert{\a= (\beta, A)} \ind\N(x_i'\beta,
A)$ &
$\mu_i \vert{y, \a} \ind\N((1-B_i)y_i + B_i x_i'\beta, V_i(1-B_i))$
\\ [2pt]
III & $\a\sim\pi(\a)$ & \\
\hline
\end{tabular*}
\end{table*}

In what follows, \mbox{Model-I} will refer to the distribution of $y \vert
\mu$ at Level-I of Table~\ref{model} which treats $\mu$ as the unknown
parameter, whereas \mbox{Model-II} will refer to the random effects model
combining \mbox{Model-I} and the Level-II distribution of $\mu\vert\a$,
which has unknown parameter $\a= (\beta, A)$. \mbox{Model-III} will refer to
the fully Bayesian model embracing all Levels I, II and III for a
single prior $\pi(\a)$ on $\a$, and is used here primarily to construct
Bayes rules to be evaluated via the assumptions of \mbox{Model-I} or \mbox{Model-II},
in the frequency sense for all $y$.

If the hyperparameter $\a= (\beta, A)$ were known in \mbox{Model-II} of
Table \ref{model}, one would use the Level-II distribution of $\mu
\vert{y,\a}$ in Table~\ref{model} to make inferences about each
component value $\mu_i$. For squared error loss the best estimator of
$\mu_i$ then would be the posterior mean, which estimates $\mu_i$ by
using the shrinkage factor
\[
B_i = \frac{V_i}{V_i + A}
\]
to compromise between the prior mean $x_i'\beta$ and the sample mean
$y_i$.

While shrinkages needn't arise for many distributions that one could
choose for Level-II, they do with the Normal distribution on $\mu_i$
because the Normal distribution on $\mu$ in Level-II is conjugate to
the Normal Level-I likelihood. Conjugate priors at Level-II lead to
linear posterior means and shrinkage coefficients for the Normal and
for other exponential family models too; see, for example, Diaconis and
Ylvisaker \cite{diaconisylvisaker79} and Morris and Lock
\cite{morrislock09}. They also are the ``$G_2$ minimax'' choice for
Level-II \cite{jacksonetal70,morris83c}.

With $k > r + 2$ components, it is not required to assume $\alpha=
(\beta, A)$ is known because information builds up through the $k$
observations $y_i$, whose distributions are governed by their shared
dependence on $(\beta, A)$ via the likelihood function given by the
right half of Level-I in Table~\ref{model}. For the rest of this
section we focus on the simplest case of the Table~\ref{model} model
with $\beta= 0$ ($r = 0$). Thus, $\a= A$ is the only unknown
hyperparameter. With equal variances, studying the case $\beta= 0$ is
much less restrictive than it might seem because use of the
orthogonality\vadjust{\goodbreak} trick described in Section~\ref{intro} allows
developments for $\beta=0$ to be extended back to the case with~$\beta$ unknown.

Early work on the equal variance case strongly emphasized \mbox{Model-I}
squared error evaluations made conditionally on $\mu$. Even so, it was
realized, for\break example, Stein \cite{stein66}, that if one also assumes
Mo-\break del-II, then it is easy to motivate shrinkage estimators and the JS
estimator, since one can estimate~$A$ by considering the likelihood of
$A$, or, equivalently, of $B_i \equiv B$. The likelihood of $B$
follows from the margi\-nal distribution of $y \vert B$ in the
inferential column of Table~\ref{model} which has the form of a Gamma
density, but conditioned on $B \le1$,
\[
L(B) = B^{k/2}\exp(-BS/2).
\]
Because of the equal variance assumption, $L(B)$ only depends on the
1-dimensional sufficient statistic\break for~$B$ in the model for $y \vert
A$:
\[
S = \sum_{i=1}^k y_i^2/V.
\]
The maximum likelihood estimate of $B$ is $\hat B = k/S$. However,
$B$ (not $A$) enters linearly in $E[\mu_i \vert y]$, and by noting
that
\[
S \vert B \sim B^{-1} \chi^2_{(k)},
\]
one sees that the James--Stein shrinkage estimate $\hat B_{\mathrm
{JS}} =
(k-2)/S$ is the best unbiased estimate of $B$. Both of these estimates
lead to shrinkage or ``empirical Bayes'' estimators of $\mu_i$ via
substituting $k/S$ or $(k-2)/S$ for the shrinkage $B$, where $B$
appears in
\[
E[\mu_i \vert{y, B}] = (1-B)y_i.
\]

Minimaxity of these and of other shrinkage estimators can be checked
via Baranchik's minimax theorem, from his 1964 dissertation
\cite{baranchik64} under Stein. Assume the equal variance Normal
setting of Table~\ref{model}, $r = 0$, $k \ge3$, and \mbox{Model-I} only.
Suppose an estimator shrinks its $k$ components toward 0 based on a
shrinkage factor of the form
\[
\hat B(S) = u(S) / S,
\]
with $u(S)$ nondecreasing and with $0 \le u(S) \le2(k-2)$. Then the
estimator is minimax for total mean\vadjust{\goodbreak} squared error risk under \mbox{Model-I},
that is, with risk at most $kV$ for all $\mu$. A similar but more
general condition for minimaxity that lets $u(S)$ be decreasing also
exists \cite{efronmorris76}. These minimaxity conditions easily extend
to include shrinkage toward a fitted $r > 0$ dimensional subspace by
making $S$ be the residual sum of squares and by accounting for the
loss of $r$ degrees of freedom, so then $0 \le u(S) \le2(k - r - 2)$
is required.

\subsection{Bayes and Formal Bayes Rules}

The model of Table~\ref{model} can be expanded to Le-\break vel-III to allow
Bayesian and formal Bayesian inferences by assuming that $\a$ in
general (in our simplified context, the unknown variance parameter $A$)
has a~proper or improper prior distribution. Shrinkage factors are
then determined as integrals over the posterior distribution of~$B$,
\[
E[B\given S] = \frac{\int_0^1 B L(B) \pi(B) \ud B}{\int_0^1 L(B)
\pi(B) \ud B}
\]
for some prior density $\pi$ on $B$.

Two obvious families of priors arise in this context, to be charted in
Figure \ref{priors}:

\begin{enumerate}
\item\textit{Scale-invariant priors on $A$}. Indexed by
constants $c \ge0$, these are improper (i.e., not
finitely integrable) formal priors, with differential elements
\[
A^{c/2} \ud A/A,\quad A > 0.
\]
As a distribution on $B$, this corresponds to
\[
B^{-c/2 - 1}(1-B)^{c/2 - 1} \ud B,\quad0 < B < 1.
\]
These have the form of Beta densities, but they do not
integrate finitely. Only propriety of the posterior
distribution is required, that is, after multiplication
by $L(B)$, which imposes the additional restriction $0 < c < k$.

\item\textit{Conjugate priors on $B$} take the form of
the likelihood function $L(B)$, but with different
values of $k, S$. We index this conjugate family by $k_0 > 2$
and by $S_0 \ge0$, perhaps thinking of them as previous
values of $k$ and $S$. Posterior propriety now requires that
$k_0$ satisfy $ k_0 + k > 0 $. The prior and posterior
densities take the same form as $L(B)$, having differential element
\[
B^{(k_0 - 2)/2}\exp(-BS_0 / 2) \ud B/B,\quad0 < B < 1.
\]
If $S_0 > 0$, these are ``truncated'' $\chi^2_{(k_0 - 2)}$
distributions on $B \le1$,\vadjust{\goodbreak} scaled by $S_0$. This second
family involves proper priors if $k_0 > 2$, known as ``Strawderman's priors''
\cite{strawderman71} when $S_0 = 0$. Strawderman showed (via
Baranchik's theorem) that the posterior mean of $B$ for these
priors provides minimax and admissible shrinkage estimators if
$k_0 \le k - 2$ (so $k \ge5$ is required). These properties
also hold if $S_0 > 0$. When $S_0 = 0$ $B$ has a
$\operatorname{Beta}((k_0 - 2)/2, 1)$ distribution and
\[
E B = (k_0 - 2)/ k_0
\]
a priori, again requiring $k_0 > 2$ for propriety.
$EB \le(k-4)/(k-2)$ is the upper limit for minimaxity,
requiring $k \ge5$. The special choice $k_0 = 4$ puts a~$\operatorname{Uniform}(0,1)$ prior distribution on $B$ and minimaxity then
requires $k \ge6$.
Derived from proper priors, the posterior mean of $\mu$, given the data
$y$ for any of these Strawderman priors, automatically qualifies as an
admissible, minimax estimator in the \mbox{Model-I} sense for quadratic loss.
\end{enumerate}

The densities of these two prior families can be combined by
multiplication (and some reparametri\-zation) to yield a 3-parameter
family with densities on~$B$ of the form
%
\begin{eqnarray}\label{conjprior}
&&p( B \given{k_0, c, S_0} )\nonumber\\
&&\quad\propto B^{(k_0 -c)/2 - 1}(1-B)^{c/2 -
1}\\
&&\qquad{}\cdot\exp(-BS_0/2) \ud B,\quad0 < B < 1.\nonumber
\end{eqnarray}
If $S_0 = 0$, this class of prior densities has the form
%
\begin{equation}\label{betaprior}
\operatorname{Beta}\bigl(\tfrac1 2 (u-k), \tfrac1 2 (k+k_0 -u)\bigr)
\end{equation}
with $u = k + k_0 - c$. They are proper only if $k_0 > c > 0$, that
is, if $ k_0 > u-k > 0$. The posterior density is proper if and only
if $k_0 + k > c > 0$ since the exponential term that also appears in
the posterior density, $\exp( -B(S + S_0)/2 )$, is bounded in $B$ so
that term cannot affect posterior propriety.

Figure \ref{priors} shows the key regions for this formal Beta
family \eqref{betaprior} of prior densities (scaled for $k=10$) in
terms of the two parameters $(u, k_0)$, ignoring the nearly irrelevant
$S_0 = 0$. It emphasizes regions when minimaxity holds, $0 \le u \le
2(k-2)$. Instead of $c$, the horizontal axis uses $u = k + k_0 - c$,
because $u$ determines minimaxity. It can be seen that as $S \to
\infty$, $S\times E[B\given S] \to u$ for these priors, and
Baranchik's theorem tells us that minimaxity for large $S$ fails unless
$0 \le u \le2(k-2)$. This condition is necessary for minimaxity, but
not sufficient.

Some explanation is in order, as follows in (a) through (h):
\begin{longlist}
\item[(a)] Priors on $B = V/(V + A)$ that lead to minimax estimators are
limited to $0 \le u \le2(k-2)$.\vadjust{\goodbreak}

\item[(b)]\hypertarget{propline} The posterior distribution is proper only if $k_0 > u - k$. The
45 degree line $k_0 = u - k$ in Figure~\ref{priors} marks the
(unattainable) lower bound for these priors. 

\item[(c)] Proper priors require $k_0 > c$, so that $u > k$. Proper priors
lie in the darkly shaded region to the right of the vertical line $u =
k$. Improper priors are those with $u \le k$.

\item[(d)]\hypertarget{scaleinv} The scale-invariant priors $A^{c/2}\,\mathrm{d}A/A$ on $A
> 0$ are on the
horizontal axis $k_0 = 0$, so $c = k - u$ in these priors. Posterior
propriety for these priors, as seen in~\hyperlink{propline}{(b)}, requires $0
\le u \le k$ so that shrinkage cannot extend all the way to the
Baranchik limit $2(k-2)$. Scale invariant priors cannot be proper.

Viewed as distributions on $\mu$, these scale-invari\-ant priors have
differential elements
\[
\ud\mu/\norm{\mu}^{u}
\]
[by integrating the $\N(0,A I_k)$ density with respect to $A^{c/2}\ud
A/A$, and using $c = k - u$]. If $u = 0$, that is, the prior located in
Figure~\ref{priors} at $(u, k_0) = (0,0)$, we have the $k$-dimensional
Lebesgue measure $\ud\mu$, which leads to using $y$ as the estimator
of $\mu$, that is, no shrinkage.\label{scaleinv}

One never should use $(u, k_0) = (k, 0)$, although researchers
sometimes make this mistake, thinking that the prior is vague because
this is Jeffreys' form $\ud A/A$ in other contexts. Actually, this
prior forces $B = 1$ a posteriori, no matter what the magnitude of~$S$
might be. Obviously this full-shrinkage estimator cannot be minimax.

\item[(e)] The conjugate priors $B^{(k_0 - 4) /2} \ud B$ on $0 < B < 1$
(setting $S_0 = 0$) form the upsloping line $k_0 = u - (k-2)$. These
have proper posteriors because this line lies above (and is parallel
to) the line $k_{0} = u - k$. They are proper if $u > k$, that is, $k_0
> 2$, being Strawderman's priors.

All these conjugate priors produce an easily calculated shrinkage
factor ($u$ need not be an integer in the Chi-squares) in this equal
variances setting:
%
\begin{equation}\label{postmean}
\hat B = E[B\given S] = \frac u S \times\frac{P[\chi^2_{(u+2)} \le S
+ S_0 ]}{P[\chi^2_{(u)} \le S + S_0]}.
\end{equation}
In this expression, $S\hat B$ is monotone increasing in~$S$ because the
ratio\vspace*{1pt} of $\chi^2_{(u+2)}$ and $\chi^2_{(u)}$ densities is monotone
increasing. Therefore, Baranchik's theorem applies and verifies minimaxity.

\item[(f)]\hypertarget{infoptim} The vertical line at $u = k-2$ denotes priors that have the
smallest \mbox{Model-I} risks as $\norm{\mu} \to\infty$. This holds because
all priors in Figure~\ref{priors} have shrinkages $E[B\given S]$ near
to $\hat B = u/S$ for large $S$, and this must occur when
$\norm{\mu}$\vadjust{\goodbreak}
is large.

On the other hand, the mean-squared-error risk for shrinkage
estimators of the form $u/S = a \hat B_{\mathrm{JS}}$, with $a =
u/(k-2)$ for any $0 \le a \le2$, is
\begin{eqnarray*}
&&E\Biggl[\sum_{i=1}^k\bigl((1-u/S)y_i - \mu_i\bigr)^2 \Big\vert
\mu\Biggr]\\
&&\quad= k -
(k-2)a(2-a)E[B_{\mathrm{JS}}].
\end{eqnarray*}
This risk is minimized uniformly at $a = 1$, showing that the
James--Stein estimator is optimal among estimators of the form $u/S$.
Combining these two facts shows that minimax priors with $u = k-2$ lead
to estimators with risk functions that, for large $\norm{\mu}$, will be
smaller than those in Figure~\ref{priors} with $u \neq k-2$.

\item[(g)] Admissibility of the resulting Bayes estimators of $\mu$ holds
immediately for proper priors, so the priors in the rightmost wedge
with $k < u \le2(k-2)$ provide admissible minimax estimators.

Improper priors may or may not produce admissible estimators. Various
estimators based on priors with $k-2 \le u \le k$ are admissible and
minimax at least if $k_0$ isn't too small. The SHP prior, which
corresponds to $(u, k_0) = (k-2, 0)$, $\ud A$ is an improper prior that
does yield an admissible estimator. More on this later.

\item[(h)] Inadmissibility holds for many (perhaps all) of the priors with
$u < k-2$. That this holds is suggested by the fact that the risk of an
estimator with $u < k-2$ can be lowered for large $\norm{\mu}$ by using
a~prior with $u = k-2$ [as argued in~\hyperlink{infoptim}{(f)} above]. Then it
seems likely that such a prior can be found on the $u = k-2$ vertical
axis of Figure~\ref{priors} that would increase shrinkage (shrinkage
generally increases in the rightward direction on Figure~\ref{priors})
with lower risk everywhere as a function of $\norm{\mu}$.
\end{longlist}

Early after it was recognized that the estimator $y$ could be uniformly
improved upon, numerous authors proposed priors captured by
Figure~\ref{priors}, motivated by Bayesian and/or admissibility
concerns. Many of these were scale-invariant priors with $k_0 = 0$,
especially with $k-2 \le u \le k-1$, for example, Stein, K. Alam, T.~Leonard, I. J. Good and D. Wallace, D. Rubin, D.~V.~Lindley and A. F. M.
Smith. Others were proposed on the conjugacy line $k_0 = u - (k-2)$,
including $\ud B/B$, that is, $(u, k_0) = (k-1, 1)$, which has
Jeffreys' form, and (being improper) falls at the edge of Strawderman's
priors. Various authors since have repeated these and other
suggestions, partly as ``reference priors.'' Our hope is that these
priors that decision theory has shown to lead to the best and most\vadjust{\goodbreak}
trustworthy estimators for the equal variances setting of
Figure \ref{priors} are ``transportable'' to the unequal
variances
setting.

%
\begin{figure*}[b]

\includegraphics{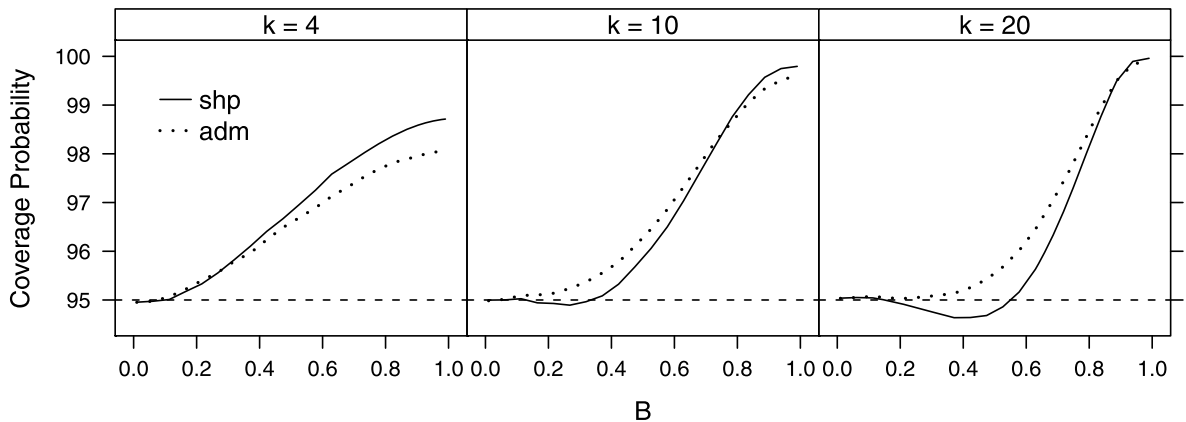}

\caption{Exact coverage probabilities against true shrinkage factor $B
= V/(V+A)$ for two equal variances rules, SHP (dark curve) and the ADM
approximation to SHP (dotted curve), with nominal 95\% coverages, for
$r = 0$ and $k = 4, 10, 20$.}\label{coverage}
\end{figure*}

Charles Stein's choice is a prior on $\mu$, not on $A$, ``Stein's
harmonic prior,'' SHP, corresponds to $\mu$ ha\-ving a measure that stems
from $c = 2$, $A$ with a flat density. It is
\[
p(\mu) \ud\mu\propto\ud\mu/ \norm{\mu}^{k-2}.
\]
By \hyperlink{scaleinv}{(d)}, this corresponds to $(u, k_0) = (k-2, 0)$ in
Figure~\ref{priors}. The term ``harmonic'' refers to the fact that the
Laplacian of the prior $\nabla^2 p(\mu)$ is uniformly equal to~0,
except at the origin where it fails to exist. Technically, since
$\nabla^2 p(0) = -\infty$, the prior is actually superharmonic
(Laplacian less than or equal to 0), a~term Stein himself employed when
showing that the resulting Bayes rule was both admissible and minimax
by \mbox{Model-I} standards \cite{stein81}. However, the term ``harmonic'' is
simpler, nearly correct, and used by most researchers.

One motivation for the SHP prior stems from an easy calculation that
shows the James--Stein shrinkage coefficient satisfies $E[B\given S] =
(k-2)/S = \hat B_{\mathrm{JS}}$\break if one assumes the (absurd) prior that
$A \sim\break
\operatorname{Uniform}[-V, \infty)$ \cite{morris83b}. Of course,
allowing $A <
0$ is illogical, and removing that part of the support for~$A$ gives $A
\sim\operatorname{Uniform}[0, \infty)$, which yields the SHP.

A second motivation is that taking $A$ uniform on $(0, \infty)$ lies
uniquely in Figure~\ref{priors} at the intersection of the
scale-invariant priors ($k_0 = 0$) and the sloped line of conjugate
priors [$k_0 = u - (k-2)$]. That is Stein's SHP. Indeed, the SHP sits
on the ``admissible boundary,'' being the scale-invariant admissible
prior that shrinks least among the admissible ones. It is also optimal
as $\norm\mu\to\infty$ ($u = k-2$). Being formal Bayes but not
proper Bayes, it provides little prior information about~$A$.\vadjust{\goodbreak} Its
conjugacy makes its shrinkages easy to compute in the equal variance
setting.

A third motivation, as will be seen, is that the aggregate conditional
posterior risk $\star R = \star R(S) < kV$ for this prior, and, in
turn, $\star R$ exceeds the unbiased estimate of the aggregate risk
$\hat R(S)$, not shown, on the SHP estimator; see Morris \cite
{morris77,morris83b}. More on this momentarily.

Using~\eqref{postmean}, the posterior mean of $B$ resulting from the
SHP prior is
\[
\hat B_{\mathrm{SHP}} = E[B \given y] = \frac{k-2} S \times\frac
{P[\chi^2_{(k)}
\le S]}{P[\chi^2_{(k-2)} \le S]}.
\]
The posterior variance of $B$ \cite{morris77,morris83b} is, for $k
\ge3$,
\begin{eqnarray*}
v &=& \var(B \given y)\\
& =& \frac2 {k-2} \hat B_{\mathrm{SPH}}^2 - (\hat B_{\mathrm
{JS}} -
\hat B_{\mathrm{SHP}})\biggl(1 - \frac{k}{k-2} \hat B_{\mathrm
{SHP}}\biggr).
\end{eqnarray*}
For the $k = 10$ hospitals we have $\hat B_{\mathrm{SHP}} = 0.668
\times0.829 =
0.571$ and $v = (0.218)^2$.

From the SHP posterior mean $\hat B_{\mathrm{SHP}}$ we obtain the
formal Bayes
rule of $\mu_i$,
\[
\hat\mu_{\mathrm{SHP},i} = E[\mu_i \given y] = (1-\hat B_{\mathrm
{SHP}})y_i.
\]

But what of interval estimates for $\mu_i$? Our Mo-\break del-III construction
via SHP suggests use of posterior probability intervals. For the SHP
these can easily be approximated after computing the posterior variance
of $\mu_i$, which for $r = 0$ is
\[
s_i^2 = \var(\mu_i \given y) = V(1 - \hat B_{\mathrm{SHP}}) + v y_i^2.
\]
Figure \ref{coverage} from Morris and Tang \cite{morristang09} shows
coverage rates of $\mu_i$ for 2-sided intervals with nominal coverage
95\%. Each interval is centered at its SHP shrinkage estimate and
approximates each of\vadjust{\goodbreak} the $k$ posterior distributions as Normally
distributed with interval widths determined by adding and subtracting
$1.96s_i$.

The true coverages in Figure \ref{coverage} for SHP are never less
than 94.5\% for any value of $A$ for any of the three values of $k = 4,
10, 20$ shown. The coverage probabilities do not depend on $i$ or on
$V$, so this is tantamount to a proof that this procedure comes close
to providing or exceeding the nominal coverage. Over-coverages rise
noticeably above 95\% as the between-groups variance $A$ approaches 0,
that is, as the shrinkage $B$ approaches~1. One must keep in mind,
however, that while these intervals are based on the posterior mean and
variance, they are not the posterior intervals because such are not
symmetric. For example, the under-coverage by 0.5\% for SHP when $k =
20$ does not account for the posterior skewness of the distributions of
$\mu_i$, which is considerable when $y_i$ is one of the extreme
observations.

Intervals based on a different estimator, determined by an
approximation technique called adjustment for density maximization
(ADM) \cite{morris88,morristang09}, also are shown in
Figure \ref{coverage}, having slightly better minimum coverage. These
estimators are described in the next section.

Componentwise intervals better than those centered at $y_i$, that is,
intervals that average being shorter than $2\times1.96\mathrm{sd}_i$,
do not exist for all $\mu$, by \mbox{Model-I} standards. Such may exist for
all $A$, when averaging over $\mu\given A$ in \mbox{Model-II}. Indeed, note
that~$s_i^2$ also can be interpreted as the Bayes risk of the SHP rule
$\hat\mu_{\mathrm{SHP},i}$,
\[
\star R_i = E[(\hat\mu_{\mathrm{SHP},i} - \mu_i)^2 \given y] = s_i^2.
\]
Let us contrast this with
\[
\hat R_i = V(1-2\hat B) + y_i^2(\hat B^2 + 2v),
\]
the unique unbiased estimate of the component risk of $\hat
\mu_{\mathrm{SHP},i}$. That is,
\[
E\hat R_i = E[(\hat\mu_{\mathrm{SHP},i} - \mu_i)^2 \given\mu]
\]
is the \mbox{Model-I} component\vspace*{1pt} risk for any value of $\mu$. Letting $\star R
= \sum_{i=1}^k s_i^2$ and $\hat R = \sum_{i=1}^k \hat R_i$, by
rearranging terms \cite{morris77,morris83} one sees that
\[
\hat R < \star R < kV.
\]
That $\star R < kV$ shows \mbox{Model-I} minimaxity of $\hat\mu_{\mathrm
{SHP}}$, since
its risk is less than that of the minimax $y$. That $\hat R < \star R$
shows that the SHP prior is so vague that its Bayes risk is more
conservative than its frequency-based unbiased estimate\vadjust{\goodbreak} of risk.
Averaging over both~$\mu$ and $y$, the $k$ componentwise risks
\[
E[\star R_i \given A] = E[(\hat\mu_i - \mu_i)^2 \given A]
\]
are all the same. Thus, each is less than $V$ for all $A \ge0$. This
establishes \mbox{Model-II} componentwise minimaxity, that is, improvement on
$y_i$ for all $A \ge0 $ and for every $i = 1, \ldots, k$.

Not only is the SHP rule componentwise minimax under \mbox{Model-II}
evaluations, but its (approximate) coverage intervals are shorter on
average than those accompanying the unbiased estimate $y$ (since $E s_i
< \sqrt V$ by Jensen's inequality). However, values of $y$ exist for
which some $s_i^2 > V$, although this happens with small probability.

For the 10 hospitals we obtain $\star R = 4.85$ and $\hat R = 3.48$.
Componentwise risks and other calculations are displayed in
Table~\ref{intereq}. Notice that some components have negative
unbiased estimates of their mean-square-error, a not uncommon
occurrence, and an undesirable feature of using this unbiased
estimation approach for assessing component risks.

%
\begin{table}
\caption{SHP estimates and posterior standard deviations of indices of
success rates in the 10 NY hospitals, and two estimates of
the~associated risk}\label{intereq}
\begin{tabular*}{\columnwidth}{@{\extracolsep{\fill}}d{2.2}d{2.2}ccd{2.3}@{}}
\hline
\multicolumn{1}{@{}l}{$\bolds{y_i}$} & \multicolumn{1}{c}{$\bolds
{\hat\mu}_{\mathrm{\mathbf{SHP}},\bolds{i}}$} & \multicolumn
{1}{c}{$\bolds{s_i}$} &
\multicolumn{1}{c}{$\bolds{\star R_i = s_i^2}$} & \multicolumn
{1}{c@{}}{$\bolds{\hat R_i}$} \\
\hline
-2.15 & -0.92 & 0.81 & 0.649 & 1.803 \\
-0.34 & -0.15 & 0.66 & 0.435 & -0.092 \\
-0.08 & -0.03 & 0.66 & 0.430 & -0.138 \\
0.01 & 0.00 & 0.66 & 0.429 & -0.141 \\
0.08 & 0.03 & 0.66 & 0.430 & -0.138 \\
0.57 & 0.24 & 0.67 & 0.445 & -0.004 \\
0.61 & 0.26 & 0.67 & 0.447 & 0.015 \\
0.86 & 0.37 & 0.68 & 0.465 & 0.170 \\
1.11 & 0.48 & 0.70 & 0.488 & 0.377 \\
2.05 & 0.88 & 0.79 & 0.629 & 1.627 \\
\hline
\end{tabular*}
\end{table}

Unfortunately, real data rarely come with equal variances, designed
experiments being the exception. Decision theorists have focused on
this symmetric case because it is simple enough to enable exact (small
sample) calculations. Decision theory has identified the SHP and other
priors close to it that lead to shrinkage estimators with good
frequency properties. Now the hope is that such priors are
``transportable'' to the unequal variances situation.

%
\begin{table*}[b]
\vspace*{-3pt}
\caption{NY hospital profiling data and shrinkages}\label{hosp}
\begin{tabular*}{\textwidth}{@{\extracolsep{\fill
}}d{2.0}d{2.2}cccccccd{2.2}c@{\hspace*{30pt}}d{2.0}d{4.0}@{\hspace*{-2pt}}}
\hline
\multicolumn{1}{@{}l}{$\bolds{i}$} & \multicolumn{1}{c}{$\bolds
{y}$} & $\mathrm{\mathbf{sd}}$ & $\bolds{\hat B}_{\mathrm{\mathbf{HB}}}$
& $\bolds{\hat B}_{\mathrm{\mathbf{F}}}$ & $\bolds{\hat
B}_{\mathrm{\mathbf{MLE}}}$ & $\bolds{\hat B}_{\mathrm
{\mathbf{ADM}}}$ & $\bolds{\hat B}_{\mathrm{\mathbf{SHP}}}$ & $\bolds{\sqrt v}$
& \multicolumn{1}{c}{$\bolds{\hat\mu}_{\mathrm{\mathbf{SHP}}}$} &
$\bolds{s}_{\mathrm{\mathbf{SHP}}}$
& \multicolumn{1}{c}{$\bolds{d}$} & \multicolumn{1}{c@{}}{$\bolds
{n}$} \\[-0.2pt]
\hline
1 & -2.07 & 2.78 & 0.079 & 0.947 & 0.952 & 0.922 & 0.926 & 0.047 &
-0.15 & 0.76 & 3 & 67 \\[-0.2pt]
2 & -0.22 & 2.76 & 0.081 & 0.946 & 0.952 & 0.921 & 0.925 & 0.047 &
-0.02 & 0.76 & 2 & 68 \\[-0.2pt]
3 & 0.58 & 1.57 & 0.249 & 0.850 & 0.864 & 0.790 & 0.808 & 0.103 & 0.11
& 0.69 & 5 & 210 \\[-0.2pt]
4 & -1.87 & 1.42 & 0.305 & 0.823 & 0.839 & 0.754 & 0.777 & 0.115 &
-0.42 & 0.70 & 11 & 256 \\[-0.2pt]
5 & -0.74 & 1.39 & 0.318 & 0.817 & 0.833 & 0.746 & 0.770 & 0.118 &
-0.17 & 0.67 & 9 & 269 \\[-0.2pt]
6 & -1.97 & 1.37 & 0.327 & 0.812 & 0.829 & 0.741 & 0.766 & 0.119 &
-0.46 & 0.70 & 12 & 274 \\[-0.2pt]
7 & -1.90 & 1.36 & 0.332 & 0.810 & 0.827 & 0.738 & 0.763 & 0.120 &
-0.45 & 0.70 & 12 & 278 \\[-0.2pt]
8 & 2.31 & 1.32 & 0.352 & 0.801 & 0.818 & 0.726 & 0.753 & 0.124 & 0.57
& 0.72 & 4 & 295 \\[-0.2pt]
9 & -0.14 & 1.22 & 0.413 & 0.774 & 0.794 & 0.694 & 0.725 & 0.133 &
-0.04 & 0.64 & 10 & 347 \\[-0.2pt]
10 & -1.21 & 1.22 & 0.413 & 0.774 & 0.794 & 0.694 & 0.725 & 0.133 &
-0.33 & 0.66 & 13 & 349 \\[-0.2pt]
11 & -1.43 & 1.20 & 0.427 & 0.769 & 0.788 & 0.687 & 0.719 & 0.134 &
-0.40 & 0.66 & 14 & 358 \\[-0.2pt]
12 & 1.56 & 1.14 & 0.473 & 0.750 & 0.770 & 0.664 & 0.700 & 0.140 &
0.47 & 0.66 & 7 & 396 \\[-0.2pt]
13 & -0.00 & 1.10 & 0.508 & 0.736 & 0.758 & 0.648 & 0.686 & 0.144 &
-0.00 & 0.62 & 12 & 431 \\[-0.2pt]
14 & 0.41 & 1.08 & 0.527 & 0.729 & 0.751 & 0.640 & 0.679 & 0.146 &
0.13 & 0.61 & 11 & 441 \\[-0.2pt]
15 & 0.08 & 1.04 & 0.568 & 0.714 & 0.736 & 0.622 & 0.664 & 0.149 &
0.03 & 0.60 & 13 & 477 \\[-0.2pt]
16 & -2.15 & 1.03 & 0.579 & 0.710 & 0.733 & 0.618 & 0.660 & 0.150 &
-0.73 & 0.68 & 22 & 484 \\[-0.2pt]
17 & -0.34 & 1.02 & 0.590 & 0.706 & 0.729 & 0.613 & 0.656 & 0.151 &
-0.12 & 0.60 & 15 & 494 \\[-0.2pt]
18 & 0.86 & 1.02 & 0.590 & 0.706 & 0.729 & 0.613 & 0.656 & 0.151 &
0.30 & 0.61 & 11 & 501 \\[-0.2pt]
19 & 0.01 & 1.01 & 0.602 & 0.702 & 0.725 & 0.608 & 0.652 & 0.152 &
0.00 & 0.60 & 14 & 505 \\[-0.2pt]
20 & 1.11 & 0.98 & 0.639 & 0.689 & 0.713 & 0.594 & 0.640 & 0.155 &
0.40 & 0.61 & 11 & 540 \\[-0.2pt]
21 & -0.08 & 0.96 & 0.666 & 0.680 & 0.704 & 0.584 & 0.631 & 0.157 &
-0.03 & 0.58 & 16 & 563 \\[-0.2pt]
22 & 0.61 & 0.93 & 0.710 & 0.666 & 0.691 & 0.568 & 0.618 & 0.160 &
0.23 & 0.58 & 14 & 593 \\[-0.2pt]
23 & 2.05 & 0.93 & 0.710 & 0.666 & 0.691 & 0.568 & 0.618 & 0.160 &
0.78 & 0.66 & 9 & 602 \\[-0.2pt]
24 & 0.57 & 0.91 & 0.742 & 0.656 & 0.681 & 0.558 & 0.609 & 0.161 &
0.22 & 0.58 & 15 & 629 \\[-0.2pt]
25 & 1.10 & 0.90 & 0.758 & 0.651 & 0.677 & 0.552 & 0.604 & 0.162 &
0.44 & 0.59 & 13 & 636 \\[-0.2pt]
26 & -2.42 & 0.84 & 0.870 & 0.619 & 0.646 & 0.518 & 0.575 & 0.167 &
-1.03 & 0.68 & 35 & 729 \\[-0.2pt]
27 & -0.38 & 0.78 & 1.000 & 0.584 & 0.611 & 0.481 & 0.542 & 0.171 &
-0.17 & 0.53 & 26 & 849 \\[-0.2pt]
28 & 0.07 & 0.75 & 1.000 & 0.565 & 0.592 & 0.461 & 0.525 & 0.173 &
0.03 & 0.52 & 25 & 914 \\[-0.2pt]
29 & 0.96 & 0.74 & 1.000 & 0.558 & 0.586 & 0.455 & 0.519 & 0.174 &
0.46 & 0.54 & 20 & 940 \\[-0.2pt]
30 & -0.21 & 0.66 & 1.000 & 0.501 & 0.529 & 0.399 & 0.469 & 0.177 &
-0.11 & 0.48 & 35 & 1193 \\[-0.2pt]
31 & 1.14 & 0.62 & 1.000 & 0.470 & 0.498 & 0.369 & 0.442 & 0.178 &
0.64 & 0.51 & 27 & 1340 \\[-0.2pt]
\hline
\end{tabular*}
\end{table*}

It should be clear that \mbox{Model-I} verifications are rarely appropriate
for scientific applications, even when equal variances obtain.
Acceptance of \mbox{Model-II}, and thus of evaluations that average over
Level-II distributions (given the hyperparameters, e.g., $A$), has
many\vadjust{\goodbreak}
advantages for applications. It makes assessing weights for the loss
function become unimportant. \mbox{Model-II} allows estimators to exist that
are minimax for every component for all $A$, not just when summed over
all components. Confidence intervals exist that are on average shorter
than standard intervals centered at $y_i$, and these also can have
(nearly) uniformly higher coverages. The unequal variance setting
gives further impetus to \mbox{Model-II} as a basis for evaluating the
operating characteristics of shrinkage procedures, and also for
constructing them from proper or improper priors that lead to good
repeated sampling properties.


\section{Approaches to Unequal Variance Data}\label{uneqvar}

In practice, equal variances are the exception ra\-ther than the rule.
The variances for all 31 NY hospi\-tals, not just the middle 10, differ
by a factor of~mo\-re than 20. Table~\ref{hosp} lists the data for these
$k = 31$ NY hospitals and several shrinkage-related estimators, to be
discussed further. The raw data contain the number of deaths $d_i$
within a month of CABG surgeries for each hospital $i$, sorted by
increasing caseload $n_i$. The indices for success rates are
calculated as
\[
y_i = C \times\bigl(\arcsin(1-2d_i/n_i) - \arcsin(1-2\bar d/ \bar
n)\bigr),
\]
a variance stabilizing transformation of the unbiased success rate
estimates $\hat p_i = d_i/n_i$, assuming Binomial data, in which case
the variance of the $y_i$ is approximately $V_i = \bar n/n_i$ (with
$\bar n = \tfrac1 k \sum_{i=1}^k n_i$). The factor~$C$ is chosen so
that the harmonic mean of the~$V_i$, that is,
\[
V_H = \frac k {\sum_{i=1}^k V_i^{-1}},
\]
is equal to 1. Larger values of $y_i$ correspond to higher success
rates. The 10 hospitals used in the\vadjust{\goodbreak} previous sections appear here as
Hospitals 15--24, but in a~different order.

The $y_i$ cannot be nearly Normally distributed\break when $n_i$ is small,
for example, Hospitals 1 and 2,
but we act here as if the $y_i$ are Normal because that distribution is
required for the estimators being considered. A more accurate model
might approximate the data $d_i$ as Poisson, as Christiansen and
Morris~\cite{christiansenmorris97} do for medical profiling. For the
remainder of this section we also focus on shrinkage to 0 ($r=0$), the
approximate average of the $y_i$.

\subsection{Minimaxity in \mbox{Model-I}}

It may seem for unequal variances that the James--Stein estimator, which
requires equal variances, can still be used. To do this, one would
divide the values $y_i$ by their standard errors $\mathrm{sd}_i =
\sqrt{V_i}$ to create equal variances and apply James--Stein to
$y_i/\mathrm{sd}_i$. Then the shrinkage $\hat B_{\mathrm{JS}} =
(k-2)/S =
0.697$, where $S = \sum y_i^2/V_i = 41.59$, emerges for estimating
$\mu_i/\mathrm{sd}_i$. Transforming back to estimate $\mu_i$ yields a
constant-shrinkage estimator
\[
\hat\mu_{\mathrm{JS},i} = (1-\hat B_{\mathrm{JS}})y_i.
\]
This procedure is \mbox{Model-I} minimax if the loss function,
\[
L(\hat\mu, \mu) = \sum_{i=1}^k W_i(\hat\mu_i - \mu_i)^2,
\]
has weights $W_i = 1/V_i = n_i/ \bar n$. However, if the loss function
has equal weights $W_i \equiv1$, then this estimator won't be minimax
when the variances, equivalently the patient case-loads $n_i$, are
substantially unequal, that is, it won't have uniformly lower mean
squared error than $y$ for all $\mu$. Does any health leader exist
with the insight to identify the proper weights~$W_i$ and the authority
to enforce their use?

For unequal variances, component shrinkages\break would be expected to depend
on $i$. How should these shrinkages be estimated, and by what
criteria should the estimates be guided? Data analysts desire more
shrinkage for larger $V_i$ and less for smaller $V_i$, a pattern
consistent with the law of large numbers, and with anticipated
regression toward the mean, both of which suggest placing greater
reliance on estimates $y_i$ that are based on more data and that have
smaller variances. Paradoxically, \mbox{Model-I} minimaxity in the unequal
variance setting requires reversed shrinkages (more shrinkage for
smaller $V_i$), as shown next.

Using an integration by parts technique pioneered by Stein
\cite{stein73} and Berger \cite{berger76}, Hudson\vadjust{\goodbreak} \cite{hudson74} and
Ber\-ger~\cite{berger76} independently developed a simple \mbox{Model-I} minimax
shrinkage estimator for the sum of (unweighted) squared errors, that
is, having risk less than $\sum V_i$, the risk of the unbiased estimate
$y$, for all $\mu$. Their estimator directly extends the James--Stein
estimator to unequal variances by shrinking each $y_i$ toward 0 using
the shrinkage factor
\[
\hat B_{\mathrm{HB},i} = \frac{(k-2)/V_i}{\sum_{j=1}^k(y_j/V_j)^2}.
\]
More generally, this estimator can be adapted easily to provide\vspace*{1pt} a
minimax estimator for any set of weights $W_i$ in the loss function (by
rescaling the $y_i$ to $W_i^{1/2} y_i$, obtaining the shrinkage factors
above, and then transforming back to the original scale). In the
special case $W_i = 1/V_i$, this rescaling will produce the James--Stein
estimator with its equal shrinkages $\hat B_{\mathrm{JS},i} \equiv\hat
B_{\mathrm{JS}}$.

With equal weights $W_i \equiv1$, the risk of this minimax estimator
has a simple unbiased estimate:
\[
\hat R_{\mathrm{HB}} = \sum_{i=1}^k V_i\bigl(1 - (k-2)\hat
B_{\mathrm{HB},i}\bigr).
\]
This is less than $\sum V_i$ for all values of $y$, because $\hat
B_{\mathrm{HB}, i} > 0$. It follows that the expectation of $\hat
R_{\mathrm{HB}}$
given $\mu$ is less than $\sum V_i$, thereby proving the Hud\-son--Berger
estimator uniformly dominates $y$ and is minimax for an equally
weighted loss function.

%
\begin{figure*}

\includegraphics{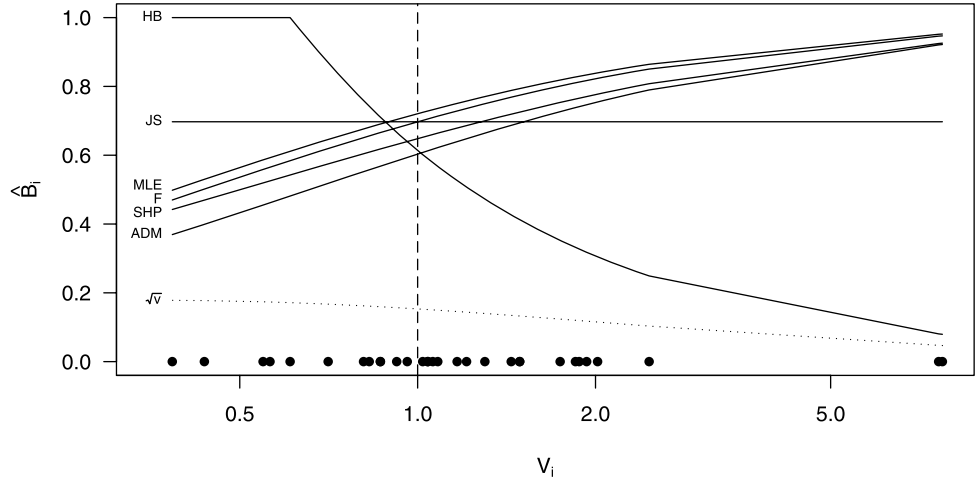}

\caption{Shrinkage factors against $V_i$ for various rules. The dots
represent $V_i$ for the 31 NY hospitals.}\label{shrinkV}
\vspace*{-1pt}
\end{figure*}

For the $k = 31$ hospitals the risk estimate of the Hudson--Berger rule
is $\hat R_{\mathrm{HB}} = 31.25$. This is 36.3\% smaller than the risk
of the unbiased estimate's\break $\sum V_i = 49.06$. Slightly more
improvement stems from using shrinkages $\min(1, \hat B_{\mathrm
{HB},i})$. Five
hospitals, Hospitals 27--31, have such $\hat B_{\mathrm{HB}, i} > 1$,
as shown in
Table~\ref{hosp}, and these shrinkages should be truncated at 1.
However, these \mbox{Model-I} minimax shrinkage factors are smallest for the
hospitals with the largest variances, even though the purpose of
combining data in these applications is to borrow strength and thereby
improve estimates for hospitals with less data.

Unfortunately, none of the 15 hospitals with the largest variances
shrinks even as much as $2/3$ of its standard error. By contrast, two of
the six hospitals already with the most data and with the smallest
variances (Hospitals 26--31) shrunk by about two of their own (small) standard
errors, a dramatic adjustment for them. This minimax estimator would
thrill the management of Hospital 26, whose negative performance
estimate $y_{26}$ (2.8 standard deviations below the mean) is shrunken
upward by 2.5 standard\vadjust{\goodbreak} deviations to make it nearly average. On the
other hand, this minimax estimator shrinks Hospitals 27--31
all the
way to 0 (the statewide average), so that Hospital 31 has its strong
positive performance $y_{31} = 1.14$ (1.84 standard deviations above
the mean) reduced by those 1.84 standard deviations so it also is
estimated as average.\vspace*{-1pt}

\subsection{Exchangeability in \mbox{Model-II}}\vspace*{-1pt}

The culprit here is the \mbox{Model-I} minimax criterion, and not the
mathematically elegant procedure derived to achieve \mbox{Model-I} minimaxity.
With substantially unequal variances and summed equally-weighted
squared error losses, achieving \mbox{Model-I} mi\-nimaxity (nearly) requires
reversed shrinkages, that is, smaller shrinkages for those components
with lar\-ger $V_i$. (``Nearly'' acknowledges that one could drastically
diminish all the larger shrinkages to eliminate the reversal, but then
with minuscule \mbox{resulting} shrinkages and no practical benefit.)
Meanwhile, procedures that do not suffer from reversed shrinkages
abound in practice, by relying instead on exchangeability assumptions
in multilevel models and on\break Bayesian and empirical Bayesian
considerations.



Figure~\ref{shrinkV} shows the Hudson--Berger \mbox{Model-I} minimax shrinkage
factors, labeled as ``HB,'' plotted against the variances $V_i$. Note
their reversed shrinkages that decrease as variances increase. The
James--Stein shrinkage factors are constant at $\hat B_{\mathrm
{JS}} =
0.697$, as
shown by the horizontal line labeled ``JS.'' Four other shrinkage
rules will be introduced next, all motivated by \mbox{Model-II}
considerations, so all with shrinkages that increase as variances
increase.

Componentwise risks and interval coverages beco\-me more valuable when
based on averages over both levels of\vadjust{\goodbreak} \mbox{Model-II}. This requires
accepting \mbox{Level-II} exchangeability for the random effects $\mu_i$ (or
when $r > 0$, accepting exchangeability of the residuals $\mu_i -
x_i'\beta$), given $A$. Shrinkages now may increase as the variances
$V_i$ increase. Exchangeability of $\mu$ (or of its residuals)
replaces assessing weights for component losses in applications. As in
the equal variances case, procedures that dominate on all $k$
components become possible, as well as confidence intervals. With
decision theoretic \mbox{Model-II} evaluations, componentwise dominance
becomes the goal.

Most data analysts and modelers of real data are familiar with
recognizing problems for which exchan\-geability assumptions are
reasonable, for example, they make such judgements routinely for error
terms when fitting regressions. Exchangeability considerations would
stop anyone from combining estimates of butterfly populations and
percentages of sports car sales to augment the estimation of the 31 NY
hospital success rates. \mbox{Model-I} standards provide no guidance on
this, in favor of requiring assessment of relative weights $W_i$ for
butterfly vs. hospital data.

With sufficiently disparate $V_i$, the minimax estimator of Hudson and
Berger is not necessarily minimax for every component by \mbox{Model-II}
evaluations. However, \mbox{Model-II} minimax shrinkage estimators do exist
for any set of $V_i$. A recent such procedure by Brown, Nie and Xie
\cite{brownetal10} produces shrinkages that increase with $V_i$ and
with componentwise squared errors smaller than $V_i$ for every $i$, for
all $A \ge0$, and for any variance pattern $V_1, \ldots, V_k$ for $k
\ge3$.



A popular \mbox{Model-II} shrinkage technique is based on the MLE of $A$. It
provides relatively simple MLE estimates of the shrinkages $\hat
B_{\mathrm{MLE},i} = V_i/(V_i + \hat A_{\mathrm{MLE}})$ and\vadjust{\goodbreak} of the
unknown means $\hat
\mu_{\mathrm{MLE},i} = (1-\hat B_{\mathrm{MLE},i})y_i$. It is often
used to construct
confidence intervals for the $\mu_i$ by estimating the conditional
variance
\[
\var(\mu_i \given{y, A}) = (1-B_i)V_i.
\]
For $r = 0$, $\hat A_{\mathrm{MLE}}$ maximizes
\[
L(A) = \sum_{i=1}^k\bigl(-S_iB_i + \log(B_i)\bigr)/2,
\]
where $S_i = y_i^2/V_i$ and $B_i = V_i/(V_i+A)$. If $r > 0$ and
Level-II in Table~\ref{model} specifies an unknown mean
\[
E[\mu_i \given\a] = x_i' \beta,
\]
then restricted maximum likelihood (REML) should be used. This can be
accomplished by analytically integrating out (not maximizing out) the
$r$-dimensio\-nal $\beta$, assuming its prior density is flat in $r$
dimensions, as in \cite{morristang09}. In this case the likelihood
$L(A)$ above would be replaced by the resulting integral over $\beta$,
and then maximization would lead to $\hat A_{\mathrm{REML}}$.

When $r > 0$, a larger value of $k$ is required for any possibility of
minimaxity, at least $k \ge3 + r$, with $k \ge5 + r$ needed for
minimaxity of the MLE in the equal variance case. The MLE shrinkages
are graphed in Figure~\ref{shrinkV} for the 31 hospitals on the curve
labeled ``MLE.''

A flaw of the MLE is that $\hat A_{\mathrm{MLE}} = 0$ occurs commonly.
This not
only dictates full shrinkage, but also when $r = 0$ the conditional
variance estimates $(1-\hat B_{\mathrm{MLE},i})V_i$ are all equal to
zero. In
such cases using these for confidence intervals asserts that $\mu_i =
0$ with 100\% confidence, a gross overstatement \cite{morristang09}.

\subsection{Construction at Level-III}

Bayesian modeling extends \mbox{Model-II} to \mbox{Model-III} by constructing
procedures from a single prior on the hyperparameters at Level-III.
Bayes and formal Bayes procedures provide posterior means, variances
and posterior distributions for the random effects $\mu_i$, given the
data. As such \mbox{Model-III} Bayesian procedures are widely used in
applications, the question is: what are their frequency properties? The
posterior moments and distributions may not be computable exactly, but
they are estimable for any particular data set and prior via MCMC and
other simulation techniques. Moreover, the fundamental theorem of
decision theory tells us that \mbox{Model-III} constructions (Bayes and formal
Bayes) are required for \mbox{Model-II} admissibility.

From the decision-theoretic perspective much mo\-re is yet to be
learned,
even for models as simple as the Normal distributions of
Table~\ref{model} in Levels I--II. It still isn't known,\vadjust{\goodbreak} even with
$r=0$, whether (formal) priors exist that provide \mbox{Model-II} minimax
estimators of $\mu$ no matter how varied the $V_i$. Beyond that, only
a little has been done in the unequal variance case to determine if
posterior probability intervals for formal priors, perhaps computed to
offer posterior coverages of 95\%, actually cover $\mu_i$ for
every~$i$, $A \ge0$ at that nominal 95\% level.

\subsubsection{Stein's prior: Transported from the equal va\-riance case}

For the family of priors discussed in the equal variances case in
Section~\ref{eqvar}, Stein's SHP stands out as the prime candidate for
minimaxity and for confidence intervals in the unequal variances
setting, assuming \mbox{Model-II} evaluations. Unfortunately, no general
theorems about these properties have been proved for the SHP, formal
mathematical proofs being hindered by the complexity of the posterior
moments and intervals. However, particular investigations with the SHP
have been encouraging.

Indeed, for any shrinkage estimator $\hat\mu_i = (1-\hat B_i)y_i$ with
$0 < \hat B_i < 1$, the difference between the component risks of $y_i$
and $\hat\mu_i$ conditioned on $A$ and $y$,
\begin{eqnarray*}
r_i & = & E[(y_i - \mu_i)^2 \given{A, y}] - E[(\hat\mu_i - \mu_i)^2
\given{A, y}] \\
& = & B_i^2y_i^2 - (B_i - \hat B_i)^2y_i^2 \\
& = & (2B_i - \hat B_i)\hat B_iy_i^2,
\end{eqnarray*}
is positive for any value of $A < V_i$, which, when integrating over
$y$, shows that the \mbox{Model-II} risk of~$\hat\mu_i$ is less than $V_i$
for any $A < V_i$. Also, SHP will dominate the unbiased estimate $y_i$
when $A$ becomes large enough, since the componentwise \mbox{Model-II} risk
converges to that of equal variances as $A$ tends to infinity.

%
\begin{figure*}

\includegraphics{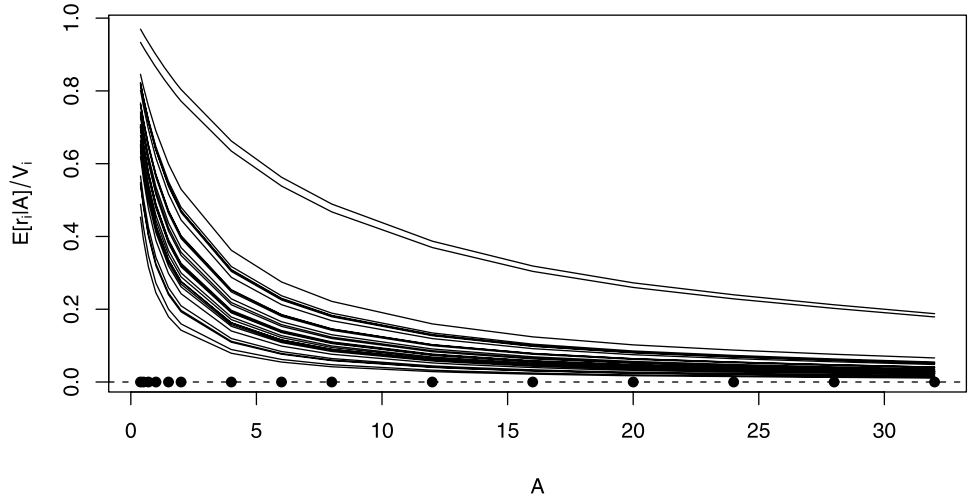}

\caption{Stochastic estimate of SHP's \mbox{Model-II} componentwise relative
risk improvement for 31 variances as in the hospitals, as a function of
$A$. The dots represent the values of $A$ at which the simulations were
performed (20,000 replicates of $y$ for each
$A$).}\label{shpdom}
\end{figure*}

The estimator $E[\mu_i \given y]$ for any prior on $A$, for each~$i$
and set of variances $V_i$, involves computing $E[B_i \given y]$. For
the SHP, with $L(A)$ being the \mbox{Model-II} likelihood of $A$, this is
\[
E[B_i \given y] = \hat B_{\mathrm{SHP},i} = \frac{\int_0^\infty
V_i/(V_i + A)
L(A) \ud A}{\int_0^\infty L(A) \ud A},
\]
and the resulting estimate of $\mu_i$ is
\[
\hat\mu_{\mathrm{SHP},i} = (1-\hat B_{\mathrm{SHP},i})y_i.
\]
As with the equal variance case, the posterior variances $s_i^2 =
\var(\mu_i \given y)$ for any prior are given by
\[
s_i^2 = V_i(1-E[B_i \given y]) + v_i y_i^2,
\]
where $v_i$ is the posterior variance of $B_i$. For SHP this is
\begin{eqnarray*}
v_i &=& \var(B_i \given y)\\
&=& \frac{\int_0^\infty V_i^2/(V_i + A)^2 L(A)
\ud A}{\int_0^\infty L(A) \ud A} - \hat B_{\mathrm{SHP},i}^2.
\end{eqnarray*}
The SHP shrinkage estimates $\hat B_{\mathrm{SHP}, i}$ for the
hospital data are
plotted in Figure~\ref{shrinkV} on the curve labeled ``SHP.'' The
associated posterior standard deviations $\sqrt{v_i}$ are given by the
dotted curve labeled ``$\sqrt v$.'' Figure~\ref{shpdom} displays a
stochastic estimate of the relative \mbox{Model-II} risk improvement of SHP
over the unbiased estimate $y_i$,
\[
\frac{V_i - E[(\hat\mu_{\mathrm{SHP},i} - \mu_i)^2 \given A]}{V_i} =
\frac{E[r_i \given A]}{V_i}
\]
for $k = 31$ and for the variance pattern of the 31 hospitals $\rv V
{31}$ as a function of $A$.

%
\begin{figure*}

\includegraphics{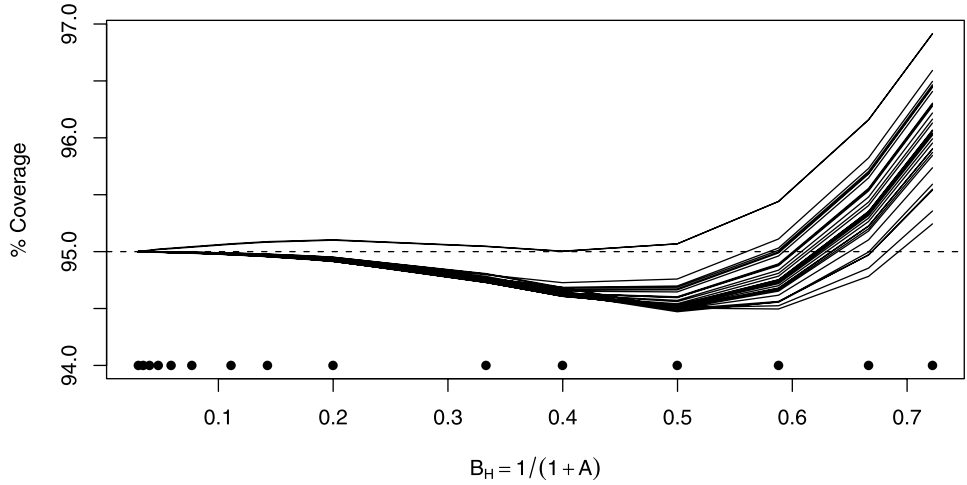}

\caption{Simulation of SHP's 95\% Normal interval \mbox{Model-II}
componentwise coverage probabilities for each of the 31 hospital
variances as a function of $B_H$. The dots represent the values of
$B_H$ at which the simulations were performed.}\label{shpcov}\vspace*{-3pt}
\end{figure*}
This was done by simulating 20,000 replicates of~$y$ at 15 different
values of $A$, and averaging the 20,000 values of $r_i/V_i$ at each
$A$. Different curves plot the risk improvement for different
components~$i$. All the curves are positive and strictly decreasing.
The cur\-ves are ordered according to their $V_i$ values, the largest
($V_1$) providing the top curve. Thus, for this variance pattern, at
least and seemingly generally, the greatest shrinkage benefit accrues
to the components with the greatest uncertainty.

%
\begin{figure*}[b]
\vspace*{-3pt}
\includegraphics{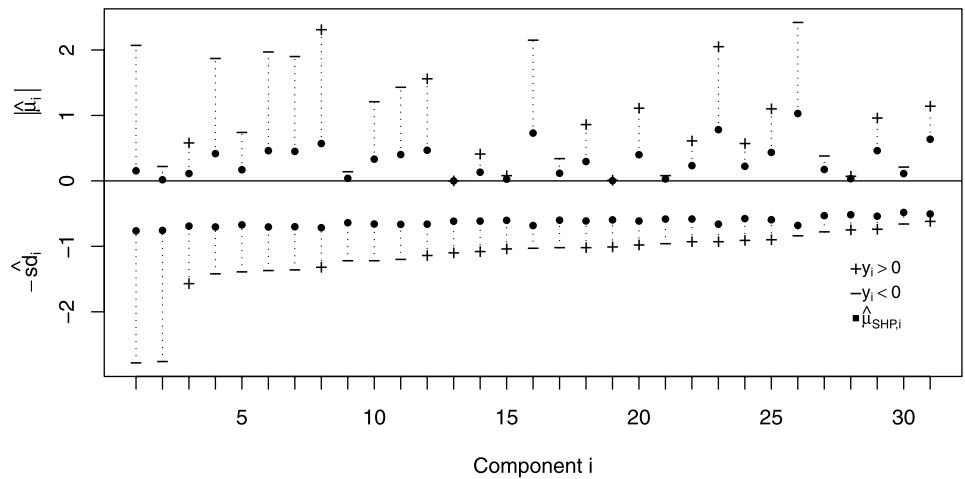}

\caption{SHP (black circles). Absolute values of unshrunken and SHP
estimates with signs indicated by ($+/-$) top half. Standard deviations
(bottom half) for SHP are always closer to 0 than $V_i$. ``Plus'' signs
indicate positive estimates $y_i \ge0$.}\label{shpvsy}
\end{figure*}

The graph's monotonicity suggests that the minimum \mbox{Model-II} risk
improvement for each component occurs as $A$ approaches infinity. That
corresponds to the limiting equal variance case. Interestingly,
despite their stochastic nature, the curves do not cross each other.
These results, although only for one data set, give hope for
establishing componentwise \mbox{Model-II} risk dominance for all $A$ of the
SHP shrinkage procedure over the unbiased estimate~$y$.\looseness=1

For equal variances, Figure~\ref{coverage} showed that $\hat
\mu_{\mathrm{SHP},i} \pm1.96s_{\mathrm{SHP},i}$ produces minimum
coverage of $\mu_i$\vadjust{\goodbreak}
very close to 95\%. Figure~\ref{shpcov} investigates the corresponding
coverage properties for the unequal variances in the pattern of the 31
NY hospitals. For each $y$ and $A$ of the previous simulation, the
coverage probability
\[
P(\mu_i \in\{\hat\mu_{\mathrm{SHP},i} \pm1.96s_{\mathrm
{SHP},i}\} \given{y,
A})
\]
of $\mu_i$ by the ``SHP Normal'' interval given $y$ and $A$ is
analytically computed from
\[
\mu_i \given{y, A} \ind\N[(1-B_i)y_i, V_i(1-B_i)],
\]
then averaged over the 20,000 values of $y$ for each~$A$.
Thus, Figure~\ref{shpcov} displays the coverage probabilities for each
Hospital $i$ using \mbox{Model-II} of Table~\ref{model} as a~function of the
harmonic mean $B_H$ of the shrinkage factors,
\[
B_H = \frac{k}{\sum_{i=1}^k B_i^{-1}} = \frac{V_H}{V_H + A} = \frac1
{1 + A},
\]
a monotone decreasing function of $A$ (recall that the 31 CABG indices
have been scaled to have $V_H = 1$).

All but two of the 31 curves exhibit a pattern similar to that of equal
variances in Figure \ref{coverage} when $k = 20$: exactly 95\% coverage
for $B_H$ close to 0, a~minimum with 0.5\% under-coverage near $B_H =
0.6$, and over-coverage for $B_H$ close to 1. The curves are
nonintersecting and increasing with $V_i$ for the 4 highest values of
$B_H$, but cross each other repeatedly for $B_H < 0.5$, presumably
because of simulation inaccuracy. The two nearly superimposed highest
curves which never (or barely) overcover $\mu_i$ correspond to the two
hospitals with the highest variances, Hospitals 1--2, these variances being
nearly 8 times the size of the 31 variances' harmonic mean. In all
cases the coverage probabilities are never below 94.5\%.\looseness=1

Figure~\ref{shpvsy} compares SHP and unshrunken estimates, and their
standard deviations for the data with the 31 NY hospitals. The
absolute value of the rules, $\abs{\hat\mu_{\mathrm{SHP},i}}$ (circle)
and $\abs{y_i}$ ($+/-$), are plotted above the $x$-axis, and the
negative standard deviations, $-s_{\mathrm{SHP},i}$ and $-\mathrm
{sd}_i$, are
plotted below. ``Plus'' signs indicate that the estimates were
positive, for example, Hospitals 3 and 8, whereas ``minus'' signs
indicate that the estimates were\vadjust{\goodbreak} negative, for example, Hospitals
1--2. It appears for these data that all the SHP coverage intervals
will be shorter than those of the unbiased estimate, although this need
not always hold for all data sets $y$, as discussed earlier for the
equal variances in Section \ref{eqvar}.\vspace*{3pt}

\subsubsection{Posterior mean versus posterior mode: The ADM technique}

Deriving the SHP rule for unequal variances requires numerical
computation of $k+1$ integrals (including the common denominator\break in
$\hat B_{\mathrm{SHP},i}$). ADM (adjustment for density maximization, Morris
\cite{morris88}) is used here for shrinkage estimation to provide a
relatively simple approximation to the SHP, as in Morris and Tang
\cite{morristang09}. To explain the ADM, the MLE provides a simple
shrinkage formula\vadjust{\goodbreak} from the mode of the likelihood $L(A)$ that is
equivalent to the posterior mode of $A$ for the SHP. However, the mode
of a right-skewed distribution like that of $A$ underestimates the
mean. Furthermore, the mean $E[B_i \given y]$ is needed, not the mode.
The ADM provides a better approximation than the MLE for shrinkage
factors while still requiring only two derivatives to approximate the
posterior distributions of $B_i \given y$. The ADM can be used with
various priors in Figure~\ref{priors}, but here we apply it to
approximate the posterior distribution of each shrinkage $B_i \given y$
when the SHP is the chosen prior distribution for $A$.

For shrinkage estimation, ADM approximates the distribution of each
$B_i = V_i/(V_i + A)$ by a Beta distribution. Because shrinkage
coefficients lie in $[0,1]$, and these coefficients linearly determine
the \mbox{Level-II} distributions (Table~\ref{model}), two-parameter Beta
distributions are the natural choice for shrinkage approximations, and
not the Normal distribution (the distribution for which MLE and the
posterior mean would coincide). When the prior on $A$ is taken to be
the SHP, and with Beta distribution approximations to $B_i = V_i/(V_i +
A)$, the ADM ``adjustment'' simply amounts to maximizing $A\cdot
L(A)$, rather than $L(A)$, for each $i$ and $V_i$. Note that\vspace*{1pt} the
maximum always occurs with $A \ge0$. Calling the maximizing value
$\hat A_{\mathrm{ADM}}$, then $E[B_i \given y]$ is approximated by
$\hat
B_{\mathrm{ADM}, i} = V_i/(V_i + \hat A_{\mathrm{ADM}})$. This ADM approach has been
used before for shrinkage estimation, for example, by\break Christiansen and
Morris \cite{christiansenmorris97}, Li and Lahiri \cite{lilahiri10}
and Morris and Tang \cite{morristang09}.

For the 31 hospitals, $\hat A_{\mathrm{ADM}} = 0.657$, so $E[B_i
\given y]$ is
approximated by $\hat B_i = V_i/(V_i + 0.657)$. The variances of $B_i$
could be obtained from the second derivative of $\log(A\cdot L(A))$ at
the adjusted mode, $\hat A_{\mathrm{ADM}} = 0.657$. The ADM shrinkages are
graphed in Figure~\ref{shrinkV} on the curve labeled ``ADM.'' They are
more conservative than those of the MLE, and indeed follow the SHP
curve closely for all but the smallest variances~$V_i$.

As was seen before in Figure~\ref{coverage}, standard errors and
interval estimates with the SHP coverages as approximated by ADM are
never perceptibly below 95\%, for equal variances and $k = 4, 10, 20$.
The ADM is readily applicable to approximate posterior point and
interval estimates for other priors on $A$ in the unequal variance
case. Further, \mbox{Model-II} evaluations of ADM include investigations by
Morris--Tang \cite{morristang09} for Normal distributions,
Everson--Morris~\cite{eversonmorris00} for multivariate\vadjust{\goodbreak} Normal data,
and Christiansen--Mor\-ris~\cite{christiansenmorris97} for Poisson data.
Evidence therein with special cases and/or with special data sets has
been quite encouraging, with no negative experiences thus far.

\subsection{Potential of the Multilevel Model: A~Useful~Rule~of~Thumb}

For equal variances, good shrinkage rules such as James--Stein or SHP
are simple enough to calculate that they can be implemented immediately
in practice. For unequal variances the calculations are much more
involved and easily accessed software may be unavailable or need to be
mastered. Researchers justifiably may ask how much they stand to gain
by fitting a hierarchical model before actually fitting it, their
alternatives being to use unbiased estimates $\hat\mu_i = y_i$ or the
fully shrunken estimates, here $\hat\mu_i = 0$ (for $r = 0$), or when
$r > 0$ to shrink all the way to a~grand mean or to a linear
regression estimate.

A helpful feature of using MLE or ADM methods to fit shrinkages,
perhaps with a model like that of Table~\ref{model}, is that a simple
point estimate $\hat A$ of $A$ suffices to estimate all shrinkage
factors $B_i$, and consequently also all means $\mu_i$. Moreover, an
estimate~$\hat A$ of $A$ leads to a simple estimate $\hat B_H$ of the
harmonic mean of the shrinkage factors $B_H$ through the identity
%
\begin{equation}\label{harmB}
B_H = V_H/(V_H + A).
\end{equation}
Analogously to its equal variance counterpart $B$, the harmonic mean
shrinkage $0 \le B_H \le1$ provides a~useful summary for gauging the
benefits of fitting a shrinkage model. Values of $B_H$ close to 0
suggest that there will be relatively little shrinkage overall, in
which case a researcher might be justified to use the unbiased
estimates $y_i$. Or, values of $B_H$ close to~1 might justify using
the fully shrunken regression estimates $x_i'b$,
\[
b = (X'V^{-1}X)^{-1}X'V^{-1}y,
\]
where $X' = [x_1, \ldots, x_k]$ and $V =
\operatorname{diag}(V_1,\ldots,V_k)$.\break Values of $B_H$ near $1/2$ give the
strongest case for estimating shrinkages.

Letting $S = \sum_{i=1}^k (y_i - x_i'b)^2/V_i$, when $r \ge0$ and the
variances are equal, $B_H = B$ and we have
\[
E[S \given A] = (k-r)/B\quad\mbox{and}\quad E[(k-r-2)/S \given A]
= B,
\]
which leads to the James--Stein estimator. When the variances are
unequal, it is easily seen for $r = 0$ that
\[
E[S \given A] = \sum_{i=1}^k (V_i+A)/V_i = k/B_H.
\]
Taken together these facts suggest a simple point estimate for $B_H$,
\[
\hat B_H = \frac{k-r-2} S = \frac{k-r-2}{k-r} \times\frac1 {\hat
\sigma^2},
\]
where
\[
\hat\sigma^2 = \frac1 {k-r} \times\sum_{i=1}^k \frac{(y_i -
x_i'b)^2}{V_i}
\]
is the mean square error from a (weighted linear) regression output.
Note that one can easily rearrange~\eqref{harmB} to solve for
\[
\hat A = V_H(1-\hat B_H)/\hat B_H.
\]
This estimate, in turn, can be used to provide simple estimates of each
individual shrinkage factors $B_i$ by $V_i/(V_i + \hat A)$. Even if
$\hat{B}_H$ is small, having this rough estimate of every $B_i$ is
useful in case there are a~few~$\hat{B_i}$ that are appreciably bigger
than 0.

These estimates of $B_i$ are plotted as the fourth and final \mbox{Model-II}
rule in Figure~\ref{shrinkV}, labeled ``F,'' giving a curve that is
almost identical to the MLE shrinkages. Data analysts can use this
easy ``rule-of-thumb'' that can be based on regression outputs for
anticipating individual and overall shrinkages, without computing more
precise shrinkage estimates. For the 31 hospitals $S = 41.59$ and
$\hat B_H = 0.697$, suggesting that a good \mbox{Model-II} rule would
outperform both the individual estimates $y_i$ and the fully shrunken
estimates, alike.


\section{Summary and Conclusions}

We have reviewed a special and relatively simple class of hierarchical
models, models for Normal distributions that have received significant
attention from a nonasymptotic (in $k$) decision-theoretic perspective.
Early equal-variance \mbox{Model-I} shrinkage estimators, evaluated by a
(unweighted) sum of squa\-red errors criterion, were found that provided
Mo\-del-I minimaxity and even admissibility. That ope\-ned exciting new
vistas. However, the great preponderance of applications (even when
Normal distributions apply) arise with unequal variances, and there
\mbox{Model-II} evaluations are seen to be much more appropriate. \mbox{Model-II}
evaluations are both less and more general than \mbox{Model-I}, less because
they average over the Level-II parameters, and more general by not
requiring judgements about appropriate weights for component losses,
and also by empowering interval estimation. A Level-II exchangeability
assumption, for example, as in Table \ref{model}, enables componentwise
\mbox{Model-II} dominance to be possible.

Many more investigations are needed in the Mo\-del-II setting for small
and moderate numbers $k$ of random effects $\mu= (\rv\mu k)'$. Does
Stein's harmonic prior (SHP) transport to the unequal variance case,
for example, by offering \mbox{Model-II} componentwise minimaxity,
conditionally on all hyperparameters, especially on all $A \ge0$? Our
experience suggests that this is entirely possible for both the equal
and the unequal variances settings, but there are no formal proofs yet.
Does the full posterior distribution, geared to offer 95\% posterior
probability of coverage for fixed data with the SHP prior, provide
intervals that cover at least 95\% of the time? Showing this with
\mbox{Model-II} would require at least 95\% coverage for every fixed value of
$A \ge0$ that holds for every component (e.g., for every hospital),
after averaging over both levels of \mbox{Model-II}. If intervals cover less
than 95\% of the cases, how close does the minimum coverage come to
95\%? How well and when do relatively simple methods for estimating
shrinkages work, like MLE and ADM methods? What \mbox{Level-III} priors lead
to \mbox{Model-II} dominance by providing componentwise minimaxity and
confidence intervals that are shorter for every component? Do SHP
intervals cover every $\mu_i$ more often for every $i, A$ than do the
standard (unshrunken) confidence intervals used by data analysts?

These theoretical questions about operating characteristics under
\mbox{Model-II} evaluations can be asked for other yet more complicated
models, especially for other distributions at Level-I and at Level-II.
Shrinkage estimators arise when fitting generalized linear multilevel
models to data that follow exponential families at Level-I, if
conjugate distributions are used for the Level-II random effects. That
is, just as Normal conjugate distributions are used at \mbox{Level-II} in
Table~\ref{model}, Gamma distributions are conjugate when Level-I
specifies Poisson likelihoods, and Betas are conjugate for Binomial
likelihoods. The advantage of conjugate distributions at Level-II is
that shrinkage factors arise in conditional means, given the
observations. Crucially, conjugate distributions are relatively
robust, having the virtue of being ``$G_2$ minimax'' among all possible
Level-II distributions (priors) in the sense of Jackson et al.
\cite{jacksonetal70,morris83c}. This helps make shrinkage
estimators simple and robust. Shrinkage factors also provide useful
summaries, so can serve a purpose like $R^2$ does with OLS regressions.

We have argued that \mbox{Model-II} and it's exchangeability assumptions are
more appropriate than Mo\-del-I for developing and evaluating shrinkage
estimators. This holds especially for applications in which
improvements would be expected to hold for every~$\mu_i$. Hospital
directors might agree to having their own hospital's performance be
estimated by combining information from other hospitals, but not unless
each was assured that doing so would make their own hospital's estimate
more accurate.

This paper argues especially that evaluations of shrinkage methods for
unequal variance data have received too little attention, relative to
the large literature on the Normal equal variances case. It is time to
change that.




\begin{thebibliography}{29}

\bibitem{baranchik64}
\begin{btechreport}[author]
\bauthor{\bsnm{Baranchik},~\bfnm{A.}\binits{A.}}
(\byear{1964}).
\btitle{Multiple regression and estimation of the mean of a multivariate normal
  population}
\btype{Technical Report 51},
\binstitution{Dept. Statistics, Stanford Univ.}
\bptok{imsref}%
\end{btechreport}
\endbibitem

\bibitem{berger76}
\begin{barticle}[mr]
\bauthor{\bsnm{Berger},~\bfnm{James~O.}\binits{J.~O.}}
(\byear{1976}).
\btitle{Admissible minimax estimation of a multivariate normal mean with
  arbitrary quadratic loss}.
\bjournal{Ann. Statist.}
\bvolume{4}
\bpages{223--226}.
\bid{issn={0090-5364}, mr={0397940}}
\bptok{imsref}%
\end{barticle}
\endbibitem

\bibitem{berger85}
\begin{bbook}[mr]
\bauthor{\bsnm{Berger},~\bfnm{James~O.}\binits{J.~O.}}
(\byear{1985}).
\btitle{Statistical Decision Theory and {B}ayesian Analysis}, \bedition{2nd
  ed.} ed.
\bpublisher{Springer}, \baddress{New York}.
\bid{mr={0804611}}
\bptok{imsref}%
\end{bbook}
\endbibitem

\bibitem{brown66}
\begin{barticle}[mr]
\bauthor{\bsnm{Brown},~\bfnm{Lawrence~David}\binits{L.~D.}}
(\byear{1966}).
\btitle{On the admissibility of invariant estimators of one or more location
  parameters}.
\bjournal{Ann. Math. Statist.}
\bvolume{37}
\bpages{1087--1136}.
\bid{issn={0003-4851}, mr={0216647}}
\bptok{imsref}%
\end{barticle}
\endbibitem

\bibitem{brown09}
\begin{bmisc}[author]
\bauthor{\bsnm{Brown},~\bfnm{L.~D.}\binits{L.~D.}}
(\byear{2009}).
\bhowpublished{Personal communication}.
\bptok{imsref}%
\end{bmisc}
\endbibitem

\bibitem{brownetal10}
\begin{barticle}[author]
\bauthor{\bsnm{Brown},~\bfnm{L.~D.}\binits{L.~D.}},
  \bauthor{\bsnm{Nie},~\bfnm{H.}\binits{H.}} \AND
  \bauthor{\bsnm{Xie},~\bfnm{X.}\binits{X.}}
(\byear{2011}).
\btitle{Ensemble minimax estimation for multivariate normal means}.
\bjournal{Ann. Statist.}
\bnote{To appear}.
\bptok{imsref}%
\end{barticle}
\endbibitem

\bibitem{christiansenmorris97}
\begin{barticle}[mr]
\bauthor{\bsnm{Christiansen},~\bfnm{Cindy~L.}\binits{C.~L.}} \AND
  \bauthor{\bsnm{Morris},~\bfnm{Carl~N.}\binits{C.~N.}}
(\byear{1997}).
\btitle{Hierarchical {P}oisson regression modeling}.
\bjournal{J. Amer. Statist. Assoc.}
\bvolume{92}
\bpages{618--632}.
\bid{issn={0162-1459}, mr={1467853}}
\bptok{imsref}%
\end{barticle}
\endbibitem

\bibitem{diaconisylvisaker79}
\begin{barticle}[mr]
\bauthor{\bsnm{Diaconis},~\bfnm{Persi}\binits{P.}} \AND
  \bauthor{\bsnm{Ylvisaker},~\bfnm{Donald}\binits{D.}}
(\byear{1979}).
\btitle{Conjugate priors for exponential families}.
\bjournal{Ann. Statist.}
\bvolume{7}
\bpages{269--281}.
\bid{issn={0090-5364}, mr={0520238}}
\bptok{imsref}%
\end{barticle}
\endbibitem

\bibitem{efronmorris75}
\begin{barticle}[author]
\bauthor{\bsnm{Efron},~\bfnm{B.}\binits{B.}} \AND
  \bauthor{\bsnm{Morris},~\bfnm{C.}\binits{C.}}
(\byear{1975}).
\btitle{Data analysis using Stein's estimator and its generalizations}.
\bjournal{J. Amer. Statist. Assoc.}
\bvolume{70}
\bpages{311--319}.
\bptok{imsref}%
\end{barticle}
\endbibitem

\bibitem{efronmorris76}
\begin{barticle}[mr]
\bauthor{\bsnm{Efron},~\bfnm{Bradley}\binits{B.}} \AND
  \bauthor{\bsnm{Morris},~\bfnm{Carl}\binits{C.}}
(\byear{1976}).
\btitle{Families of minimax estimators of the mean of a multivariate normal
  distribution}.
\bjournal{Ann. Statist.}
\bvolume{4}
\bpages{11--21}.
\bid{issn={0090-5364}, mr={0403001}}
\bptok{imsref}%
\end{barticle}
\endbibitem

\bibitem{eversonmorris00}
\begin{barticle}[mr]
\bauthor{\bsnm{Everson},~\bfnm{Philip~J.}\binits{P.~J.}} \AND
  \bauthor{\bsnm{Morris},~\bfnm{Carl~N.}\binits{C.~N.}}
(\byear{2000}).
\btitle{Inference for multivariate normal hierarchical models}.
\bjournal{J. R. Stat. Soc. Ser. B Stat. Methodol.}
\bvolume{62}
\bpages{399--412}.
\bid{doi={10.1111/1467-9868.00239}, issn={1369-7412}, mr={1749547}}
\bptok{imsref}%
\end{barticle}
\endbibitem

\bibitem{hudson74}
\begin{btechreport}[author]
\bauthor{\bsnm{Hudson},~\bfnm{H.~M.}\binits{H.~M.}}
(\byear{1974}).
\btitle{Empirical Bayes estimation}
\btype{Technical Report 58},
\binstitution{Dept. Statistics, Stanford Univ.}
\bptok{imsref}%
\end{btechreport}
\endbibitem

\bibitem{jacksonetal70}
\begin{barticle}[mr]
\bauthor{\bsnm{Jackson},~\bfnm{D.~A.}\binits{D.~A.}},
  \bauthor{\bsnm{O'Donovan},~\bfnm{T.~M.}\binits{T.~M.}},
  \bauthor{\bsnm{Zimmer},~\bfnm{W.~J.}\binits{W.~J.}} \AND
  \bauthor{\bsnm{Deely},~\bfnm{J.~J.}\binits{J.~J.}}
(\byear{1970}).
\btitle{{$\mathcal{G}\sb{2}$}-minimax estimators in the exponential family}.
\bjournal{Biometrika}
\bvolume{57}
\bpages{439--443}.
\bid{issn={0006-3444}, mr={0270491}}
\bptok{imsref}%
\end{barticle}
\endbibitem

\bibitem{jamesstein61}
\begin{bincollection}[mr]
\bauthor{\bsnm{James},~\bfnm{W.}\binits{W.}} \AND
  \bauthor{\bsnm{Stein},~\bfnm{Charles}\binits{C.}}
(\byear{1961}).
\btitle{Estimation with quadratic loss}.
In \bbooktitle{Proc. 4th Berkeley Sympos. Math. Statist. Probab.}
\bvolume{I}
\bpages{361--379}.
\bpublisher{Univ. California Press}, \baddress{Berkeley, CA}.
\bid{mr={0133191}}
\bptok{imsref}%
\end{bincollection}
\endbibitem

\bibitem{lilahiri10}
\begin{barticle}[mr]
\bauthor{\bsnm{Li},~\bfnm{Huilin}\binits{H.}} \AND
  \bauthor{\bsnm{Lahiri},~\bfnm{P.}\binits{P.}}
(\byear{2010}).
\btitle{An adjusted maximum likelihood method for solving small area estimation
  problems}.
\bjournal{J.~Multivariate Anal.}
\bvolume{101}
\bpages{882--892}.
\bid{doi={10.1016/j.jmva.2009.10.009}, issn={0047-259X}, mr={2584906}}
\bptok{imsref}%
\end{barticle}
\endbibitem

\bibitem{morris77}
\begin{barticle}[author]
\bauthor{\bsnm{Morris},~\bfnm{C.}\binits{C.}}
(\byear{1977}).
\btitle{Interval estimation for empirical Bayes generalizations of Stein's
  estimator.
The Rand Paper Series, The Rand Corporation}.
\bptok{imsref}%
\end{barticle}
\endbibitem


\bibitem{morris83}
\begin{barticle}[mr]
\bauthor{\bsnm{Morris},~\bfnm{Carl~N.}\binits{C.~N.}}
(\byear{1983}).
\btitle{Parametric empirical {B}ayes inference: Theory and applications (with
  discussion)}.
\bjournal{J. Amer. Statist. Assoc.}
\bvolume{78}
\bpages{47--65}.
\bid{issn={0162-1459}, mr={0696849}}
\bptnote{check related}%
\bptok{imsref}%
\end{barticle}
\endbibitem


\bibitem{morris83c}
\begin{barticle}[mr]
  \bauthor{\bsnm{Morris},~\bfnm{Carl~N.}\binits{C.~N.}} (\byear{1983}).
  \btitle{Natural exponential families with quadratic variance functions:
  Statistical theory}.
  \bjournal{Ann. Statist.}
  \bvolume{11}
  \bpages{515--529}.
  \bid{issn={0090-5364}, mr={0696064}}
  \bptok{imsref}
  \end{barticle}
  \endbibitem



\bibitem{morris83b}
\begin{bincollection}[mr]
\bauthor{\bsnm{Morris},~\bfnm{Carl~N.}\binits{C.~N.}}
(\byear{1983}).
\btitle{Parametric empirical {B}ayes confidence intervals}.
In \bbooktitle{Scientific Inference, Data Analysis, and Robustness ({M}adison,
  {W}is., 1981)}.
\bseries{Publ. Math. Res. Center Univ. Wisconsin}
\bvolume{48}
\bpages{25--50}.
\bpublisher{Academic Press}, \baddress{Orlando, FL}.
\bid{mr={0772762}}
\bptok{imsref}%
\end{bincollection}
\endbibitem


\bibitem{morris88}
\begin{bincollection}[mr]
\bauthor{\bsnm{Morris},~\bfnm{C.~N.}\binits{C.~N.}}
(\byear{1988}).
\btitle{Approximating posterior distributions and posterior moments}.
In \bbooktitle{Bayesian Statistics 3 ({V}alencia, 1987)}
\bpages{327--344}.
\bpublisher{Oxford Univ. Press}, \baddress{New York}.
\bid{mr={1008054}}
\bptok{imsref}%
\end{bincollection}
\endbibitem


\bibitem{morrislock09}
\begin{barticle}[mr]
\bauthor{\bsnm{Morris},~\bfnm{Carl~N.}\binits{C.~N.}} \AND
  \bauthor{\bsnm{Lock},~\bfnm{Kari~F.}\binits{K.~F.}}
(\byear{2009}).
\btitle{Unifying the named natural exponential families and their relatives}.
\bjournal{Amer. Statist.}
\bvolume{63}
\bpages{247--253}.
\bid{doi={10.1198/tast.2009.08145}, issn={0003-1305}, mr={2750349}}
\bptok{imsref}%
\end{barticle}
\endbibitem




\bibitem{morristang09}
\begin{barticle}[author]
\bauthor{\bsnm{Morris},~\bfnm{C.}\binits{C.}} \AND
\bauthor{\bsnm{Tang},~\bfnm{R.}\binits{R.}}
(\byear{2011}).
\btitle{Estimating random effects via adjustment for density maximization}.
\bjournal{Statist. Sci.}
\bvolume{26}
\bpages{271--287}.
\bptok{imsref}%
\end{barticle}
\endbibitem



\bibitem{stein55}
\begin{binproceedings}[mr]
\bauthor{\bsnm{Stein},~\bfnm{Charles}\binits{C.}}
(\byear{1956}).
\btitle{Inadmissibility of the usual estimator for the mean of a multivariate
  normal distribution}.
In \bbooktitle{Proc. 3rd {B}erkeley {S}ympos. {M}ath. {S}tatist. {P}robab.}
\bvolume{I}
\bpages{197--206}.
\bpublisher{Univ. California Press}, \baddress{Berkeley}.
\bid{mr={0084922}}
\bptok{imsref}%
\end{binproceedings}
\endbibitem

\bibitem{stein66}
\begin{bincollection}[mr]
\bauthor{\bsnm{Stein},~\bfnm{Charles}\binits{C.}}
(\byear{1966}).
\btitle{An approach to the recovery of inter-block information in balanced
  incomplete block designs}.
In \bbooktitle{Research {P}apers in {S}tatistics ({F}estschrift {J}. {N}eyman)}
\bpages{351--366}.
\bpublisher{Wiley, London}.
\bid{mr={0210232}}
\bptok{imsref}%
\end{bincollection}
\endbibitem

\bibitem{stein73}
\begin{binproceedings}[mr]
\bauthor{\bsnm{Stein},~\bfnm{Charles}\binits{C.}}
(\byear{1974}).
\btitle{Estimation of the mean of a~multivariate normal distribution}.
In \bbooktitle{Proceedings of {P}rague {S}ymposium on {A}symptotic {S}tatistics
  ({C}harles {U}niv., {P}rague, 1973)}
\bvolume{II}
\bpages{345--381}.
\bpublisher{Charles Univ.}, \baddress{Prague}.
\bid{mr={0381062}}
\bptok{imsref}%
\end{binproceedings}
\endbibitem

\bibitem{stein62}
\begin{barticle}[mr]
\bauthor{\bsnm{Stein},~\bfnm{C.~M.}\binits{C.~M.}}
(\byear{1962}).
\btitle{Confidence sets for the mean of a multivariate normal distribution.}
\bjournal{J. Roy. Statist. Soc. Ser. B}
\bvolume{24}
\bpages{265--296}.
\bid{issn={0035-9246}, mr={0148184}}
\bptok{imsref}%
\end{barticle}
\endbibitem

\bibitem{stein81}
\begin{barticle}[mr]
\bauthor{\bsnm{Stein},~\bfnm{Charles~M.}\binits{C.~M.}}
(\byear{1981}).
\btitle{Estimation of the mean of a multivariate normal distribution}.
\bjournal{Ann. Statist.}
\bvolume{9}
\bpages{1135--1151}.
\bid{issn={0090-5364}, mr={0630098}}
\bptok{imsref}%
\end{barticle}
\endbibitem

\bibitem{strawderman71}
\begin{barticle}[mr]
  \bauthor{\bsnm{Strawderman},~\bfnm{William~E.}\binits{W.~E.}}
  (\byear{1971}).
  \btitle{Proper {B}ayes minimax estimators of the multivariate normal mean}.
  \bjournal{Ann. Math. Statist.} \bvolume{42} \bpages{385--388}.
  \bid{issn={0003-4851}, mr={0397939}}
  \bptok{imsref}
  \end{barticle}
\endbibitem\vspace*{-2pt}

\end{thebibliography}
\end{document}